\documentclass[journal,twoside,comsoc]{IEEEtran}
\usepackage{graphicx,cite,amssymb,amsmath,bm}
\usepackage{mathrsfs}
\usepackage[pagewise]{lineno}
\usepackage{color,soul}
\usepackage{tabulary}
\usepackage{multirow}
\usepackage{cases}
\usepackage{stfloats}
\usepackage{url}
\usepackage{booktabs}
\usepackage{algpseudocode}
\usepackage{algorithm,algpseudocode}
\usepackage{algorithmicx}

\ifCLASSOPTIONcompsoc
    \usepackage[caption=false, font=normalsize, labelfont=sf, textfont=sf, subrefformat=parens, labelformat=parens]{subfig}
\else
\usepackage[caption=false, font=footnotesize, subrefformat=parens, labelformat=parens]{subfig}
\fi

\usepackage{amsmath,amssymb}
\usepackage{mathtools} 
\usepackage{xspace}

\newcommand{\Secref}[1]{Section~\ref{#1}}
\newcommand{\Figref}[1]{Fig.~\ref{#1}}

\newcommand{\herm}{^\text{H}}

\newcommand{\trans}{^\text{T}}

\newcommand{\bx}{\mathbf{x}}

\newcommand{\bI}{\mathbf{I}}

\newcommand{\bA}{\mathbf{A}}

\newcommand{\bb}{\mathbf{b}}
\newcommand{\bc}{\mathbf{c}}

\newcommand{\bv}{\mathbf{v}}

\newcommand{\CN}{\mathcal{CN}}
\newcommand{\SNR}{\textsc{snr}\xspace}

\newcommand{\SINR}{\textsc{sinr}\xspace}

\newcommand{\MMSE}{\textsc{mmse}\xspace}

\newcommand{\boundellipse}[3]
{(#1) ellipse [x radius=#2,y radius=#3]
}

\newtheorem{proposition}{Proposition}

\newtheorem{remark}{Remark}

\DeclareMathOperator{\var}{\mathsf{Var}}

\DeclareMathOperator*{\argmin}{arg\,min}

\newcommand{\EX}[1]{\mathsf{E}\left\{{#1}\right\}}

\newcommand{\varx}[1]{\var\left\{{#1}\right\}}

\newcommand{\cov}[1]{\mathsf{Cov}\left\{{#1}\right\}}

\newcommand{\C}{\mathbb{C}}

\newcommand{\Pd}{\rho_{\mathrm{d}}}
\newcommand{\Pdp}{\rho_{\mathrm{d,p}}}
\newcommand{\Pup}{\rho_{\mathrm{u,p}}}
\newcommand{\wtdk}{\tilde{w}_{\mathrm{d},k}}
\newcommand{\wdk}{w_{\mathrm{d},k}}

\newcommand{\tauc}{\tau_\mathrm{c}}
\newcommand{\taup}{\tau_\mathrm{u,p}}
\newcommand{\taud}{\tau_\mathrm{d,d}}
\newcommand{\taudp}{\tau_\mathrm{d,p}}
\newcommand{\tauu}{\tau_\mathrm{u,d}}

\newcommand{\norm}[1]{{ \left\Vert #1 \right\Vert }}

\makeatletter
\def\@setsize#1#2#3#4{
    \@nomath#1
    \let\@currsize#1
    \baselineskip #2
    \baselineskip \baselinestretch\baselineskip
    \parskip \baselinestretch\parskip
    \setbox\strutbox \hbox{
        \vrule height.7\baselineskip
            depth.3\baselineskip
            width\z@}
    \skip\footins \baselinestretch\skip\footins
    \normalbaselineskip\baselineskip#3#4}
\makeatother

\makeatletter
\newcommand{\setstretch}[1]{
    \def\baselinestretch{#1}%
    \@currsize
    }
\makeatother



\def\BibTeX{{\rm B\kern-.05em{\sc i\kern-.025em b}\kern-.08em
    T\kern-.1667em\lower.7ex\hbox{E}\kern-.125emX}}

\newcounter{eqcnt1}
\newcounter{eqcnt2}
\newcounter{eqcnt3}
\newcounter{eqcnt4}

\algnewcommand{\IfThen}[2]{
  \State \algorithmicif\ #1\ \algorithmicthen\ #2}

\begin{document}
\begin{figure*}[t!]
\normalsize
This paper was submitted for publication in IEEE Transactions on Wireless Communications on September 6, 2018.  It was finally accepted for publication on July 22, 2019 and published on August 14, 2019. DOI: 10.1109/TWC.2019.2933831. 

\

\textcopyright~2019 IEEE. Personal use of this material is permitted.  Permission from IEEE must be obtained for all other uses, in any current or future media, including reprinting/republishing this material for advertising or promotional purposes, creating new collective works, for resale or redistribution to servers or lists, or reuse of any copyrighted component of this work in other works.
\vspace{20cm}
\end{figure*}

\newpage

\title{Downlink Training in Cell-Free Massive MIMO: \\A Blessing in Disguise}
\author{Giovanni~Interdonato,~\IEEEmembership{Student Member,~IEEE,} Hien~Quoc~Ngo,~\IEEEmembership{Member,~IEEE,} P{\aa}l~Frenger,~\IEEEmembership{Member,~IEEE,} and~Erik~G.~Larsson,~\IEEEmembership{Fellow,~IEEE}%
\thanks{This paper was supported by the European Union's Horizon 2020 research
and innovation programme under grant agreement No~641985 (5Gwireless), and the Swedish Research Council (VR). The work of H.~Q.~Ngo was supported by the UK Research and Innovation Future Leaders Fellowships under Grant MR/S017666/1. Part of this work was presented at the 2016 IEEE Global Communications Conference (GLOBECOM)~\cite{interdonato2016dlpilot}.}%
\thanks{G.~Interdonato and E.~G.~Larsson are with the Department of Electrical Engineering (ISY), Link\"{o}ping University, 581 83 Link\"{o}ping, Sweden (e-mail: giovanni.interdonato@liu.se; erik.g.larsson@liu.se).}%
\thanks{H.~Q.~Ngo is with Queen's University Belfast, Belfast BT7 1NN, UK (e-mail: hien.ngo@qub.ac.uk).}%
\thanks{P.~Frenger is with Ericsson Research, Ericsson AB, 581 12 Link\"{o}ping, Sweden (e-mail: pal.frenger@ericsson.com). This work was conducted when G.~Interdonato was with Ericsson Research, Ericsson AB.}%
}

\markboth{}%
{Interdonato \MakeLowercase{\textit{et al.}}: Downlink Training in Cell-Free Massive MIMO: A Blessing in Disguise}

\maketitle

\begin{abstract}
Cell-free Massive MIMO (multiple-input multiple-output) refers to a distributed Massive MIMO system where all the access points (APs) cooperate to coherently serve all the user equipments (UEs), suppress inter-cell interference and mitigate the multiuser interference.
Recent works demonstrated that, unlike co-located Massive MIMO, the \textit{channel hardening} is, in general, less pronounced in cell-free Massive MIMO, thus there is much to benefit from estimating the downlink channel. 

In this study, we investigate the gain introduced by the downlink beamforming training, extending the previously proposed analysis to non-orthogonal uplink and downlink pilots. 
Assuming single-antenna APs, conjugate beamforming and independent Rayleigh fading channel, we derive a closed-form expression for the per-user achievable downlink rate that addresses channel estimation errors and pilot contamination both at the AP and UE side. 
The performance evaluation includes max-min fairness power control, greedy pilot assignment methods, and a comparison between achievable rates obtained from different capacity-bounding techniques. 
Numerical results show that downlink beamforming training, although increases pilot overhead and introduces additional pilot contamination, improves significantly the achievable downlink rate. 
Even for large number of APs, it is not fully efficient for the UE relying on the statistical channel state information for data decoding.
\end{abstract}


\section{Introduction}
\label{sec:intro}
\IEEEPARstart{I}{n cell-free} Massive MIMO (multiple-input multiple-output)~\cite{Nayebi2015a,Ngo2017b} 
 a very large number of geographically distributed access points (APs) coherently\footnote{Coherent transmission requires accurate synchronization among the APs. A possible implementation is described in~\cite{Interdonato2018a}.} serve a
  smaller number of user equipments (UEs), in the same time-frequency resources. 
  The APs cooperate, by being connected to  a central processing unit (CPU). 
  Each UE experiences no cell boundaries as it is surrounded by serving APs, hence the term \textit{cell-free}. 

By combining the benefits from the time-division duplex (TDD) Massive MIMO concept, the distributed architecture and the signal co-processing at multiple APs, cell-free Massive MIMO guarantees ubiquitous communications at higher spectral efficiency thanks to the additional macro-diversity and a greater ability to control the interference. Cell-free Massive MIMO is the ultimate embodiment of concepts as network MIMO~\cite{Venkatesan2007a},  multi-cell MIMO cooperative network~\cite{Gesbert2010a}, coordinated multi-point with joint transmission (CoMP-JT)~\cite{Boldi2011a}, and virtual MIMO~\cite{Hong2013b}.

In the canonical form of cell-free Massive MIMO, every AP participates in the service of every UE.
However, with any reasonable power control policy, the result is that effectively, only the APs that are geographically close to a given UE will participate in its service. The result is ``user-centric'' transmission---a concept also known as ``user-specific dynamic clustering'' from the MIMO cooperative networks literature~\cite{Bjornson2013d, Garcia2010a} and CoMP~\cite{Baracca2012b}.

Moreover, leveraging the channel reciprocity of  TDD operation, precoding can be conveniently designed by using    channel estimates acquired via uplink pilots. Therefore the channel estimation overhead is independent of the number of APs. Since channel estimation and precoding can be performed locally at each AP, cell-free Massive MIMO constitutes a scalable way to implement the network MIMO concept.

In co-located Massive MIMO, the UEs do not need to estimate the downlink channel as data decoding  relying on   long-term statistical CSI at the UE is efficient~\cite{Marzetta2016a}, by virtue of   \textit{channel hardening}. The term channel hardening is used to describe a fading channel that behaves almost deterministically~\cite{Hochwald2004a}, that is, the instantaneous channel gain tends to its mean value, after coherent combining.
This phenomenon is a direct consequence of the \textit{law of the large numbers}, and it is observed at the receiver when a signal is transmitted by a large number of antennas over multiple independently fading channels. Essentially, the channel fluctuations averaged over the antennas, and their impact becomes smaller as more antennas are added.    
Hence, despite the channel randomness, the channel hardens because of the increased spatial diversity. 

In light of that, the channel hardening phenomenon is certainly present in Massive MIMO systems. Although channel hardening is not necessary for Massive MIMO to work, it is beneficial for the following reasons: $(i)$ it can alleviate the need for downlink pilots~\cite{Ngo2017a}, since UEs can reliably decode data relying only on statistical CSI; $(ii)$ it makes 
the standard ``use-and-forget'' capacity lower bounds of \cite{Marzetta2016a} more tight; 
$(iii)$ it simplifies resource allocation, as this  can be carried out on the large-scale fading time scale \cite{Bjornson2016b}, and $(iv)$ it improves reliability, since the channel is nearly deterministic. 

Conversely, in cell-free Massive MIMO with low and moderate network density, the channel vectors depend only on a small number of multipath components, as a given UE observes only a few dominant contributions from its closest APs. 
Hence, even though  the number of APs is large,   the channel hardening phenomenon is
less pronounced than in cellular  Massive MIMO. 
In preliminary work \cite{interdonato2016dlpilot}, we indirectly concluded this by observing
 that cell-free Massive MIMO benefits much more from using downlink pilots than co-located Massive MIMO. 
 Later,  an investigation of the channel hardening phenomenon in cell-free Massive MIMO using a stochastic geometry approach was  provided in~\cite{Chen2018b}. 
 Our analysis in~\cite{interdonato2016dlpilot} was constrained to orthogonal uplink and downlink pilots, and did not provide  achievable downlink rate expressions in closed form.

In this work, we consider a cell-free Massive MIMO system with single-antenna APs, conjugate beamforming and downlink beamforming training. The technical contributions are:
\begin{itemize}
\setlength\itemsep{.5em}
\item We derive a closed-form expression for an (approximate) achievable rate of the downlink channel with finite number of APs and UEs, independent Rayleigh fading, and beamformed downlink pilots. This expression  accounts for estimation errors and pilot contamination both at the AP and UE side.
It is a generalization of the result in~\cite{interdonato2016dlpilot}.  
\item We provide a tight upper bound on the   achievable rate and use this to  formulate an optimization problem for   max-min fairness power control. 
\item We devise a \textit{sequential convex approximation} (SCA) algorithm to globally solve the power control optimization problem, and show that very few iterations are needed for the  algorithm to converge. Moreover, since the contribution from the downlink training is involved in the problem formulation, the resulting rates  are significantly higher than the rates achieved by employing the power control optimization given in \cite{Ngo2017b}. This power control policy requires global knowledge of the large-scale fading (statistical CSI) at the CPU and its computation  is performed at the large-scale (slow) fading time scale.
\item We propose a greedy algorithm for uplink and downlink pilot assignment. This extends the algorithm proposed in~\cite{Ngo2017b} by jointly selecting the uplink and downlink pilot pair, for each UE, that maximizes the smallest UE rate.
\item We quantitatively compare the performance provided by cell-free Massive MIMO with downlink training with the case when the UEs only have access to   statistical CSI, and to
the lower capacity bound for non-coherent detection given in \cite{Caire2017a}. 
\item We investigate the  downlink training gain for different pilot training durations, shadow fading models, and power control policies.
\end{itemize}

\section{System Model} \label{Sec:SysModel}

We analyze a cell-free Massive MIMO system operating in TDD mode. Let $M$ be the number of APs that coherently serve $K$ active UEs, with $M>K$, in the same time-frequency resources. Both APs and UEs are herein assumed to be equipped with a single antenna. The APs are deployed in a wide area without boundaries in a random or well-planned fashion, while the UEs are uniformly randomly placed. A fronthaul network connects all the APs with a CPU, which is responsible for collection and distribution of payload data, downlink power control and pilot assignment.  

The precoding scheme we consider in this study is conjugate beamforming, also known as maximum-ratio transmission. Although it does not represent the optimal precoder, performing such   linear processing offers low operational complexity with inexpensive hardware components. In addition, unlike zero-forcing, conjugate beamforming does not require channel state information (CSI) sharing among APs and CPU, which reduces  the fronthaul network load. Therefore, CSI acquisition and precoding can be carried out locally at each AP simply by leveraging the channel reciprocity of   a TDD system.   

We consider a standard block-fading channel model which incorporates both small-scale and large-scale fading. Let $g_{mk} = \sqrt{\beta_{mk}}h_{mk}
$ be the channel response between the $k$th UE and the $m$th AP,
where $h_{mk}$ represents the small-scale fading, and $\beta_{mk}$ is the large-scale fading. 
The small-scale fading coefficients $\{h_{mk}\}$ are independent identically distributed (i.i.d.) random variables (RVs), $h_{mk} \sim \mathcal{CN}(0,1)$, for $m = 1,\ldots,M$, $k = 1, \ldots, K$. The large-scale fading includes path-loss and shadowing, and the coefficients $\{\beta_{mk}\}$ are constant over multiple coherence intervals. Hence, we assume that $\{\beta_{mk}\}$ coefficients are estimated a priori and known whenever required.

The TDD coherence interval is $\tauc$ samples long, and consists of four phases:
$(i)$ uplink training, $(ii)$ uplink data transmission, $(iii)$ downlink training, and $(iv)$ downlink data transmission. Let $\taup$, $\taudp$ be the number of samples per coherence interval spent for the transmission of uplink and downlink pilots, respectively. We indicate with $\tauu$, $\taud$ the number of samples per coherence interval spent on  the transmission of uplink and downlink data, respectively. The length of the coherence interval is given by $\tauc = \taup + \tauu + \taudp + \taud$.

In the uplink training phase, all the UEs synchronously send proper symbol sequences, referred to as \textit{pilots}, to the APs. These uplink pilots, known a-priori at both the ends of the link, enable the APs to estimate the uplink channel from different UEs. These estimates are used by the APs both for uplink detection and, by leveraging the channel reciprocity property, for downlink precoding. Downlink pilots allow UEs to estimate its effective channel towards each AP. Based on the downlink estimates, the UEs can reliably decode downlink data. Since in this work we focus on the downlink performance, the analysis on the uplink transmission phase is omitted. 

\subsection{Uplink Training and Channel Estimation}
\label{sec:UPtraining}

During the uplink training, all the UEs synchronously send their own pilot sequences to all the APs. Each AP needs to estimate the channel once in every coherence interval.

Let $\sqrt{\taup}\bm{\varphi}_k \in \C^{\taup}$ be the pilot sequence sent by the $k$th UE, $k=1,...,K$, where $\norm{\bm\varphi_k}^2 =1$. We assume that any two pilot sequences are either identical or mutually orthonormal, that is 
\begin{equation}
\label{eq:uplink-pilot-design}
\bm\varphi_k\herm \bm\varphi_{k^\prime} = \left\{ \begin{tabular}{ll}
1, & if $\bm\varphi_k = \bm\varphi_{k^\prime}$, \\
0, & otherwise.
\end{tabular}
\right.
\end{equation}
The $m$th AP receives a $\taup\times1$ vector, which is a linear superposition of $K$ pilots given by
\begin{equation}
\label{eq:uplinkpilot}
\textbf{y}_{\mathrm{up},m} = \sqrt{\taup \Pup} \sum\nolimits^K_{k=1} g_{mk} \bm{\varphi}_k + \textbf{w}_{\mathrm{up},m},
\end{equation}
where $\Pup$ is the normalized transmit signal-to-noise ratio (\SNR) of the uplink pilot symbol, and $\textbf{w}_{\mathrm{up},m} \! \in \! \C^{\taup}$ is the additive noise. The elements of $\textbf{w}_{\mathrm{up},m}$ are i.i.d. $\mathcal{CN}(0,1)$ RVs. 

Channel estimation is carried out locally and autonomously by each AP. More specifically, in order to estimate the channel $g_{mk}$, the $m$th AP processes the received pilot vector by projecting it onto the known pilot sequence $\bm{\varphi}\herm_k$, as follows
\begin{align}
\label{eq:uplinkpilotprojection}
\check{y}_{\mathrm{up},mk} &= \bm{\varphi}\herm_k\textbf{y}_{\mathrm{up},m} \nonumber \\
&=\sqrt{\taup\Pup} g_{mk} \!+\!\sqrt{\taup\Pup}\sum\limits^K_{k'\neq k} g_{mk'} \bm{\varphi}\herm_k \bm{\varphi}_{k'}\!+\!\tilde{w}_{mk},
\end{align}
where $\tilde{w}_{mk} \triangleq \bm{\varphi}\herm_k\textbf{w}_{\mathrm{up},m}\sim \mathcal{CN}(0,1)$. The second term in \eqref{eq:uplinkpilotprojection} represents the uplink pilot contamination effect. 
The uplink pilot design in~\eqref{eq:uplink-pilot-design} ensures that $\check{y}_{\mathrm{up},mk}$ is a sufficient statistic, and estimates based on $\check{y}_{\mathrm{up},mk}$ are optimal~\cite{Ngo2017b}.

Given $\check{y}_{\mathrm{up},mk}$, the $m$th AP performs   linear \textit{minimum mean-square error} (\MMSE) estimation of the channel $g_{mk}$ as follows 
\begin{equation}
\label{eq:mmse}
\hat{g}_{mk} = \frac{\EX{\check{y}^*_{\mathrm{up},mk}g_{mk}}}{\EX{\vert\check{y}_{\mathrm{up},mk}\vert^2}}\check{y}_{\mathrm{up},mk} = c_{mk}\check{y}_{\mathrm{up},mk},
\end{equation}
where
\begin{equation}
\label{eq:cmk}
c_{mk} \triangleq \frac{\sqrt{\taup\Pup}\beta_{mk}}{\taup\Pup \sum^K_{k'=1}\beta_{mk'}|\bm{\varphi}\herm_k \bm{\varphi}_{k'}|^2+1}.
\end{equation}
The channel estimation error is given by $\tilde{g}_{mk} \triangleq g_{mk} - \hat{g}_{mk}$. By definition, $\hat{g}_{mk}$ and $\tilde{g}_{mk}$ are uncorrelated, owing on the linear \MMSE properties~\cite{Kay1993a}. Furthermore,
the estimate and estimation error are jointly Gaussian distributed, thus they are statistically independent. 

The mean-square of the estimated channel $\hat{g}_{mk}$ is denoted by $\gamma_{mk}$ and given by
\begin{equation}
\label{eq:defGamma}
\gamma_{mk} \triangleq \EX{|\hat{g}_{mk}|^2} = \sqrt{\taup\Pup} \beta_{mk} c_{mk}.
\end{equation}
The channel estimate and the estimation error are distributed as $\hat{g}_{mk} \sim \CN(0,\gamma_{mk})$ and $\tilde{g}_{mk} \sim \CN(0,\beta_{mk}-\gamma_{mk})$, respectively.
\begin{remark}
The variance of the estimated channel $\gamma_{mk}$ also measures the quality of the estimation process. In fact, $\beta_{mk} \geq~\gamma_{mk}$, with equality if the estimation is error-free.    
\end{remark}

\subsection{Downlink Data Transmission}
In the downlink data transmission phase, the APs use the channel estimates to properly define the precoders. With conjugate beamforming, the precoder consists of the conjugate of the channel estimate. Therefore, the data signal transmitted by the $m$th AP to all the UEs is   
\begin{equation}
\label{eq:APtransmittedsignal}
x_m = \sqrt{\Pd}\sum\nolimits^K_{k=1} \sqrt{\eta_{mk}} \ \hat{g}^*_{mk} q_k,
\end{equation}
where $q_k$ is the data symbol intended for the $k$th UE, $\EX{|{q_k}^2|}=1$. The symbols $\{q_k\}$ have zero mean and unit variance, and they are uncorrelated. The normalized transmit \SNR related to the data symbol is denoted by $\Pd$. Lastly, $\eta_{mk}$, $m=1,...,M$, $k=1,...,K$, are the power control coefficients satisfying the following average power constraint at each AP: 
\begin{equation}
\label{eq:pwConstraint}
\EX{|x_m|^2}\leq\Pd.
\end{equation}
Substituting (\ref{eq:APtransmittedsignal}) into (\ref{eq:pwConstraint}), the power constraint can be rewritten as
\begin{equation}
\label{eq:pwConstraintGamma}
\sum\nolimits_{k=1}^K \eta_{mk} \gamma_{mk} \leq 1, \quad \forall m.
\end{equation}    
The $k$th UE receives a linear combination of the signals transmitted by all the APs given by
\begin{align}
\label{eq:receiveddownlinksignal}
r_{\mathrm{d},k} &= \sum\nolimits^M_{m=1} g_{mk} x_m + \wdk \nonumber \\
&=  \underbrace{\sqrt{\Pd} a_{kk} q_{k}}_{\text{desired signal}} + \underbrace{\sqrt{\Pd} \sum\nolimits^K_{k' \neq k} a_{kk'} q_{k'}}_{\text{inter-user interference}} + \underbrace{\wdk}_{\text{noise}}
\end{align}
where
\begin{align}\label{eq:akkdef}
a_{kk'} \triangleq \sum\nolimits^M_{m=1} \sqrt{\eta_{mk'}} {g}_{mk}
\hat{g}^*_{mk'}, ~ k'=1,...,K,
\end{align}
describes the effective channel gain. 
The noise at the receiver, denoted by $\wdk$, is $\mathcal{CN}(0,1)$. The $k$th UE must have a sufficient knowledge of $a_{kk}$ in order to reliably decode $q_k$. There are, at least, four approaches to decoding at the UE:
\begin{enumerate}
\item Rely on hardening, and assume that $a_{kk} \approx \EX{a_{kk}}$. This is the ``canonical'' approach in the Massive MIMO literature~\cite{Marzetta2016a}.
This is non-preferred in cell-free Massive MIMO because of the lack of hardening. 
\item Use a blind algorithm to explicitly estimate $a_{kk}$. This idea was developed for cellular Massive MIMO in~\cite{Ngo2017a}.
Whether this idea could be extended to cell-free Massive MIMO is an open question.
\item Perform non-coherent decoding, that does not rely on an explicit estimate of $a_{kk}$. A capacity lower bound for this scheme is given in~\cite{Caire2017a}, and shown in~\eqref{eq:caire-lower-bound}. However, a practical decoding method is yet to be developed.
\item Use downlink pilots to explicitly estimate $a_{kk}$.
\end{enumerate}

\subsection{Downlink Training and Channel Estimation} \label{sec:DLtraining}

In this study, we assume that the downlink pilots are beamformed to the UEs by using conjugate beamforming, as in~\cite{interdonato2016dlpilot}. 
A similar training scheme, but for conventional Massive MIMO, was used in~\cite{Ngo2013b}. 
Such beamforming of pilots is used in many systems in practice (e.g., \textit{demodulation reference signals}, DM-RS, in LTE~\cite{Dahlman2011a}).
This scheme has the advantage to be scalable in that its channel estimation overhead is independent of the number of APs, but rather scales with the number of UEs. It is also a fully distributed scheme in that precoding can be performed by each AP independently, by using only local CSI.

Let $\sqrt{\taudp}\bm{\psi}_{k} \in \mathbb{C}^{\taudp}$, $k=1,...,K$, be the downlink pilot sequence intended for UE $k$, $\Vert\bm{\psi}_{k}\Vert^2=1$, and $\taudp$ be the downlink pilot length. Similar to the uplink case, we assume that any two downlink pilot sequences are either identical or mutually orthonormal, that is 
\begin{equation}
\label{eq:downlink-pilot-design}
\bm\psi_k\herm \bm\psi_{k^\prime} = \left\{ \begin{tabular}{ll}
1, & if $\bm\psi_k = \bm\psi_{k^\prime}$, \\
0, & otherwise.
\end{tabular}
\right.
\end{equation} 
The $\taudp\times1$ downlink pilot vector transmitted by the $m$th AP is given by
\begin{equation}
\label{eq:DLpilot} \bm{x}_{m,\mathrm{p}} =
\sqrt{\taudp\Pdp}\sum\limits^K_{k=1}
\sqrt{\eta_{mk}} \hat{g}^*_{mk} \bm\psi_{k},
\end{equation}
where $\Pdp$ is the normalized transmit \SNR related to the downlink pilot symbol. 
The power each AP spends on downlink pilots per coherence interval is 
\begin{align}
\label{eq:pilot-transmit-power}
&\EX{\norm{\bm{x}_{m,\mathrm{p}}}^2} = \taudp\Pdp \EX{\norm{\sum\limits^K_{k=1}
\sqrt{\eta_{mk}} \hat{g}^*_{mk} \bm\psi_{k}}^2} \nonumber \\
&\quad=\taudp\Pdp \sum_{k = 1}^K \eta_{mk} \gamma_{mk} \nonumber \\
&\quad\quad+ \taudp\Pdp \sum_{k = 1}^K \sum^K_{k^\prime \neq k} \sqrt{\eta_{mk}\eta_{mk^\prime}} \bm\psi\herm_{k} \bm\psi_{k^\prime} \EX{\hat{g}_{mk}\hat{g}^*_{mk^\prime}}.
\end{align}
Due to uplink pilot contamination, $\EX{\hat{g}_{mk}\hat{g}^*_{mk^\prime}} \neq 0$ only if $\bm\varphi_{k^\prime} = \bm\varphi_k$. 
Constraining the transmit power by imposing a  bound  on \eqref{eq:pilot-transmit-power}
leads to a lengthy analytical expression that considerably complicates the subsequent
analysis.  To address this issue, we constrain the pilot assignment such that, for any pair of UEs $k$ and $k^\prime$, with $k \neq k^\prime$, it holds
\begin{equation}
\label{eq:pilot-assignment-constraint}
\bm\psi\herm_{k} \bm\psi_{k^\prime} = 0, \text{ if } \bm\varphi_{k^\prime} = \bm\varphi_k,  
\end{equation}
that is, orthogonal downlink pilots are assigned to those UEs that use identical uplink pilots.    
Under the constraint~\eqref{eq:pilot-assignment-constraint}, the second term in~\eqref{eq:pilot-transmit-power} is zero:
\begin{equation}
\label{eq:cross-terms-zero}
\taudp\Pdp \sum_{k = 1}^K \sum^K_{k^\prime \neq k} \sqrt{\eta_{mk}\eta_{mk^\prime}} \bm\psi\herm_k \bm\psi_{k^\prime} \EX{\hat{g}_{mk}\hat{g}^*_{mk^\prime}} = 0.
\end{equation}
Consequently, constraining 
$\EX{\norm{\bm{x}_{m,\mathrm{p}}}^2} \leq \taudp\Pdp$
is equivalent to 
$\sum_{k = 1}^K \eta_{mk} \gamma_{mk}  \leq 1,$
which in turn has the same form as the data power constraint~\eqref{eq:pwConstraintGamma}.
Note that the constraint on the pilot assignment in~\eqref{eq:pilot-assignment-constraint} is
imposed here only for analytical convenience.  While this constraint does limit the
freedom in the pilot assignment, as will be discussed in Section~\ref{subsec:pilot-assignment}, this limitation is not significant in cases of practical interest.
In Section~\ref{subsec:pilot-assignment}, we describe a joint uplink and downlink pilot assignment that satisfies~\eqref{eq:pilot-assignment-constraint}.

The corresponding $\taudp\times1$ downlink pilot vector received by the $k$th UE is given by  
\begin{equation}
\label{eq:receivedDLpilot}
\textbf{y}_{\mathrm{dp},k} = \sqrt{\taudp\Pdp} \sum\limits^K_{k'=1} a_{kk'} \bm\psi_{k'} + \textbf{w}_{\mathrm{dp},k},
\end{equation}
where $\textbf{w}_{\mathrm{dp},k}$ is a receiver noise vector, whose elements are i.i.d. $\mathcal{CN}(0,1)$ RVs. 

In order to estimate the effective downlink channel $a_{kk}$, the $k$th UE processes the received downlink pilot vector by projecting it onto the known downlink pilot sequences $\bm{\psi}\herm_k$ as follows
\begin{align}
\label{eq:y_k}
\check{y}_{\mathrm{dp},k}\! &=\!\bm{\psi}\herm_{k} \textbf{y}_{\mathrm{dp},k} \nonumber \\
&=\!\sqrt{\taudp\Pdp} a_{kk} \!+\!\sqrt{\taudp\Pdp} \sum\limits^K_{k' \neq k} a_{kk'} \bm\psi\herm_{k} \bm\psi_{k'}\!+\!n_{\mathrm{p},k},
\end{align}
where $n_{\mathrm{p},k} \triangleq \bm{\psi}\herm_k\textbf{w}_{\mathrm{dp},k} \sim \mathcal{CN}(0,1)$. The second term in~\eqref{eq:y_k} represents the downlink pilot contamination effect.
The downlink pilot design in~\eqref{eq:downlink-pilot-design} ensures that $\check{y}_{\mathrm{dp},k}$ is a sufficient statistic, and estimates based on $\check{y}_{\mathrm{dp},k}$ are optimal. 

Given $\check{y}_{\mathrm{dp},k}$, $k$th UE performs the linear \MMSE estimation of $a_{kk}$ as, according to~\cite{Kay1993a},
\begin{align}
\label{eq:a_kk}
\hat{a}_{kk} &= \EX{a_{kk}} + \frac{\cov{a_{kk},\check{y}_{\mathrm{dp},k}}}{\cov{\check{y}_{\mathrm{dp},k},\check{y}_{\mathrm{dp},k}}}(\check{y}_{\mathrm{dp},k} - \EX{\check{y}_{\mathrm{dp},k}}).
\end{align}

\begin{proposition} \label{prop:MMSE-DL-estimation}
The linear \MMSE estimate of the effective downlink channel gain at the $k$th UE, denoted by $\hat{a}_{kk}$, under independent Rayleigh fading channel and conjugate beamforming precoding scheme, is given by~\eqref{eq:a_kk} where
\begin{align}
&\EX{a_{kk}} = \sum\nolimits_{m=1}^M \sqrt{\eta_{mk}} \gamma_{mk}, \label{eq:Ea_kk} \\
&\cov{a_{kk},\check{y}_{\mathrm{dp},k}} = \sqrt{\taudp\Pdp} \sum\nolimits^M_{m=1} \eta_{mk} \gamma_{mk} \beta_{mk},  \label{eq:cov-akk-yk} \\
&\cov{\check{y}_{\mathrm{dp},k},\check{y}_{\mathrm{dp},k}}\!=\!1\!+\!\taudp\Pdp \sum\limits^M_{m=1} \sum\limits^K_{k'=1} \eta_{mk'} \gamma_{mk'} \beta_{mk} |\bm\psi\herm_{k} \bm\psi_{k'}|^2, \label{eq:cov-yk-yk} \\
&\text{and} \nonumber \\
&\EX{\check{y}_{\mathrm{dp},k}} = \sqrt{\taudp\Pdp} \sum\nolimits_{m=1}^M  \sqrt{\eta_{mk}} \gamma_{mk}. \label{eq:Ey_dpk}
\end{align}
\begin{IEEEproof}
See Appendix~\ref{app:MMSE-DL-proof}. 
\end{IEEEproof}
\end{proposition}
\begin{remark}
Consider the special case in which all pilot sequences are orthogonal; then the channel estimate in~\eqref{eq:a_kk} is identical to the one defined in~\cite{interdonato2016dlpilot}.
\end{remark}
The channel estimation error $\tilde{a}_{kk}$ is given by $\tilde{a}_{kk} = a_{kk} - \hat{a}_{kk}$. The channel estimate and the estimation error are uncorrelated, but not independent. 

\section{Performance Analysis} \label{sec:performance-analysis}

\subsection{Approximate Achievable Downlink Rate}

By following the same methodology as in~\cite{interdonato2016dlpilot}, we propose a closed-form expression for an approximate achievable rate that takes into account the channel estimation errors and the pilot contamination both at the AP and UE side.

\begin{figure}[!t]
\centering
\includegraphics[width=.9\columnwidth]{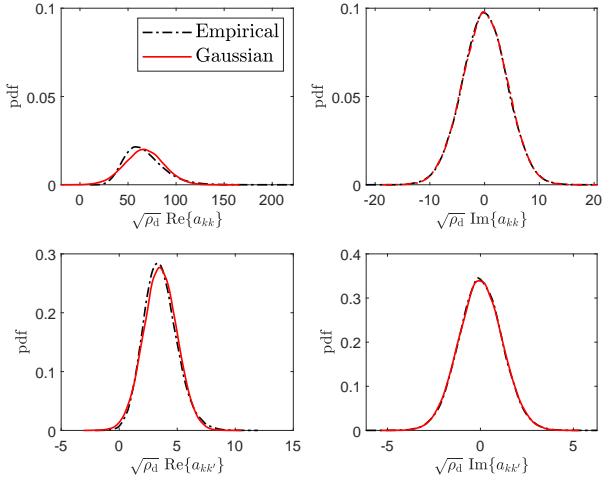}
\caption{The Gaussian and the empirical pdfs of
  $a_{kk}$ and $a_{kk'}$, $\forall k^\prime\!\neq\!k$. Here, $M\!=\!100$, $K\!=\!2$ and $\taup\!=\!1$. The $\{\beta_{mk}\}$ coefficients are modeled as in~\cite{Ngo2017b}.}
\label{fig:pdfs}
\end{figure}

A downlink achievable rate, in case of imperfect CSI at the receiver, can be obtained by using the capacity-bounding technique for fading channel with non-Gaussian noise and \textit{side information} as in~\cite[Sec. 2.3.5]{Marzetta2016a} and~\cite{Medard2000a}.  
Specifically, the received signal at UE $k$, expressed in~\eqref{eq:receiveddownlinksignal}, can be written as
\begin{align}
\label{eq:r_dk}
r_{\mathrm{d},k} = \sqrt{\Pd}\ {a}_{kk} q_{k} + \wtdk,
\end{align}
where $ \wtdk \triangleq \sqrt{\Pd}\ \sum^K_{k' \neq k} a_{kk'} q_{k'} + \wdk $ is the effective non-Gaussian noise. We assume that $q_{k^\prime}$ has zero mean and is independent of $a_{k{k^\prime}}$, for all $k$ and $k^\prime$.
Then we have
\[ \EX{\wtdk\! \mathrel{\big|}\! \hat{a}_{kk}}\!=\!\EX{q_k^\ast\wtdk\! \mathrel{\big|} \!\hat{a}_{kk}}\!=\!\EX{a_{kk}^\ast q_k^\ast\wtdk\! \mathrel{\big|}\! \hat{a}_{kk}}\! =\! 0. \]
The corresponding achievable rate is~\cite[Sec. 2.3.5]{Marzetta2016a}
\begin{equation}
\label{eq:capacity-side-information}
R_k\! \geq\! \EX{\!\log_2\!\left(\!1\!+\!\frac{\Pd \left| \EX{a_{kk} \mathrel{\big|} \hat{a}_{kk}} \right|^2}{\!\Pd\! \sum\limits^K_{k'=1}\! \EX{\!{|a_{kk'}|^2 \!\mathrel{\big|}\! \hat{a}_{kk}}\!}\!-\!\Pd \left| \EX{a_{kk} \!\mathrel{\big|}\! \hat{a}_{kk}\!} \right|^2\! +\!1\!}  \right)},
\end{equation}
where the outer expectation is taken with respect to the downlink channel estimate $\hat{a}_{kk}$. By applying the Cram\'{e}r central limit theorem\footnote{Cram\'{e}r central limit theorem~\cite{billingsley2008}: Let $X_1, X_2, ..., X_n$ be independent circularly symmetric complex RVs. Assume that $X_i$ has zero mean and variance $\sigma^2_i$. If $s^2_n = \sum^n_{i=1} \sigma^2_i \rightarrow \infty$ and $\sigma_i/s_n \rightarrow 0$, as $n\rightarrow \infty$, then $\frac{\sum^n_{i=1} X_i}{s_n} \xrightarrow{d} \mathcal{CN}(0,1), \ \text{as } n \rightarrow \infty$. Given AP $m$, $\hat{g}_{mk'}$ is function of $g_{mi}$,  $i = 1,\ldots,K$, and of the noise $\tilde{w}_{mk'}$; see~\eqref{eq:mmse}. If we define $\hat{g}_{mk'}~=~f(g_{mi},\tilde{w}_{mk'})$, then $g_{mk} f(g_{mi},\tilde{w}_{mk'})$ and $g_{nk} f(g_{ni},\tilde{w}_{nk'})$, $\forall n \neq m$, are independent because $g_{mi}$ and $g_{ni}$ are independent by assumption $\forall n \neq m$, so are $\tilde{w}_{mk'}$ and $\tilde{w}_{nk'}$.}, we have
\begin{align}
\label{eq:approxAk1}
a_{kk'}\!&=\!\sum\nolimits^M_{m=1}\!\sqrt{\eta_{mk'}} g_{mk} \hat{g}^\ast_{mk'}\! \nonumber \\ &\xrightarrow{d}\! \CN\!\left(\bm{\varphi}\herm_{k} \bm{\varphi}_{k'}\!\sum\nolimits_{m=1}^M\!\sqrt{\eta_{mk'}} \gamma_{mk'} \frac{\beta_{mk}}{\beta_{mk'}}, \varsigma_{kk'}\! \right)\!, \nonumber \\
&\text{ as } M \rightarrow \infty, k^\prime \neq k, \\
\label{eq:approxAkk}
a_{kk}\!&=\!\sum\nolimits^M_{m=1} \sqrt{\eta_{mk}} g_{mk} \hat{g}^\ast_{mk} \nonumber \\ 
&\xrightarrow{d}\!\CN\!\left(\sum\nolimits_{m=1}^M \sqrt{\eta_{mk}} \gamma_{mk}, \varsigma_{kk} \right)\!,\text{ as } M \rightarrow \infty,
\end{align}
where $\varsigma_{kk'} \triangleq \sum_{m=1}^M \eta_{mk'} \beta_{mk}
\gamma_{mk'}$ is the variance of the effective downlink channel, and $\xrightarrow{d}$ denotes convergence in distribution. Since we are considering cell-free Massive MIMO, where $M$ is large, $a_{kk^\prime}$ and $a_{kk}$ can be approximated by Gaussian RVs distributed as the right-hand side of~\eqref{eq:approxAk1} and ~\eqref{eq:approxAkk}, respectively. These approximations \eqref{eq:approxAk1} and \eqref{eq:approxAkk} are corroborated by  the numerical results shown in~\Figref{fig:pdfs}. 
We observe that the empirical probability density functions (pdfs) of the effective downlink channel gain, assuming $M = 100$, almost overlap the pdfs of the corresponding Gaussian RVs.  
Hence, the effective downlink channel gains can be safely considered as Gaussian RVs, even for finite and relatively small $M$.  

\begin{remark}
According to~\eqref{eq:approxAkk} and~\Figref{fig:pdfs}, the expected phase of $a_{kk}$ is close to zero. Hence, the use of constant envelope modulation schemes (e.g., PSK) would make the estimation of $a_{kk}$ unnecessary. However, amplitude information will be required in practice, for example, in order to properly normalize log-likelihood ratios fed to the channel decoder. Hence, unless fully non-coherent decoding could be performed, estimates of $a_{kk}$ are desirable.
\end{remark}
 
The approximations \eqref{eq:approxAk1} and \eqref{eq:approxAkk} allow us to reduce the capacity lower bounds in~\eqref{eq:capacity-side-information} to the following approximate achievable downlink rate\footnote{We stress that, although \eqref{eq:DLrateApprox} is a very good approximation, this expression is not rigorously correct capacity lower bound, as $\{a_{kk^\prime}\}~\forall k$ are non-Gaussian in general.}
\begin{equation}
\label{eq:DLrateApprox} R_k\! \approx\! \EX{\!\log_2 \!\left( \!1 \!+\!
\frac{\Pd |\hat{a}_{kk}|^2}{\Pd \EX{|\tilde{a}_{kk}|^2} \!+\! \Pd \!\sum\limits^K_{k' \neq k} \!\EX{|a_{kk'}|^2\! \mathrel{\big|}\! \hat{a}_{kk}} \!+ \!1} \!\right)
\!}.
\end{equation}
Expression~\eqref{eq:DLrateApprox}, although easier to compute than~\eqref{eq:capacity-side-information}, is still tricky due to the presence of the conditional expectation and, more importantly, is not in closed form.  Closed-form achievable rate expressions are desirable when working with system optimization, power control and resource allocation. Driven by these reasons, we further approximate~\eqref{eq:DLrateApprox} by using the   fact that
\begin{equation}
\label{eq:approx-2}
\EX{\log_2 \left(1+\frac{X}{Y}\right)} \approx \log_2 \left(1+\frac{\EX{X}}{\EX{Y}}\right),
\end{equation}
if $X$ and $Y$ are both sums of nonnegative RVs.
This approximation
does not require $X$ and $Y$ to be independent~\cite{Zhang2014a}, and it 
becomes more and more accurate as the numbers of RVs in the sums that define  $X$ and $Y$ increase. 
By applying~\eqref{eq:approx-2} to~\eqref{eq:DLrateApprox}, we obtain the following approximation for the achievable downlink rate:
\begin{equation}
\label{eq:DLrateApprox2} 
R_k\!\approx R_k^{\mathsf{cf}}\!=\!\log_2 \!\left( \!1 \!+\!
\frac{\Pd\EX{|\hat{a}_{kk}|^2}}{\Pd\EX{|\tilde{a}_{kk}|^2}\!+\!\Pd\!\sum\limits^K_{k' \neq k} \EX{|a_{kk'}|^2}\!+\!1} \!\right).
\end{equation}
\begin{proposition} \label{prop:DLrateCF}
A closed-form expression for an approximate achievable downlink rate of the transmission from the APs to the $k$th UE in a cell-free Massive MIMO system with conjugate beamforming, non-orthogonal uplink and downlink pilots\footnote{The achievable rate expression under the assumption of mutually orthogonal uplink and downlink pilots, namely $\taup = \taudp = K$, was given in~\cite{interdonato2016dlpilot}.}, for any finite $M$ and $K$, is given by~\eqref{eq:DLrateCF} at the top of the next page, where $\kappa_k = \varx{\hat{a}_{kk}}$, that is \addtocounter{equation}{1}
\begin{align}
\label{eq:kappa-def}
\kappa_k &= \frac{\taudp\Pdp \left(\sum\nolimits^M_{m=1} \eta_{mk} \gamma_{mk} \beta_{mk} \right)^2}{1 + \taudp\Pdp \sum\nolimits^M_{m=1} \sum\nolimits^K_{k'=1} \eta_{mk'} \gamma_{mk'} \beta_{mk} |\bm\psi\herm_{k} \bm\psi_{k'}|^2} \nonumber \\
&= \frac{\taudp\Pdp \varsigma_{kk}^2}{1 + \taudp\Pdp \sum\nolimits^K_{k'=1} \varsigma_{kk^\prime} |\bm\psi\herm_{k} \bm\psi_{k'}|^2}. 
\end{align}		
\begin{IEEEproof}
See Appendix~\ref{app:DL-rate-proof}. 
\end{IEEEproof}		
\end{proposition}

\begin{figure*}[!t]
\normalsize
\setcounter{eqcnt1}{\value{equation}}
\setcounter{equation}{30}
\begin{equation}
\label{eq:DLrateCF} R_k^{\mathsf{cf}} = \log_2 \!\left( \!1 \!+\!
\frac{\Pd\left( \sum\nolimits_{m=1}^M \sqrt{\eta_{mk}} \gamma_{mk} \right)^2+ \Pd\kappa_k}{\Pd(\varsigma_{kk}-\kappa_k) + \Pd \sum\limits_{k' \neq k}^K \left[ \varsigma_{kk^\prime} + \left|\bm{\varphi}_{k'}\herm\bm{\varphi}_{k}\right|^2 \left(\sum\limits_{m=1}^M \sqrt{\eta_{mk'}} \gamma_{mk'} \frac{\beta_{mk}}{\beta_{mk'}} \right)^2 \right] + 1} \!\right).
\end{equation}
\setcounter{equation}{\value{eqcnt1}}
\hrulefill
\vspace*{4pt}
\end{figure*}

The numerator of the effective \SINR (signal-to-interference-plus-noise ratio) is called \textit{coherent beamforming gain} and reflects the power of the desired signal. The first term in the denominator represents the variance of the downlink channel estimation error. Clearly, the better the \MMSE estimate is, the smaller the variance of the estimation error is. The term $\kappa_k$ is the variance of the downlink channel estimate, which includes the effects of the downlink pilot contamination. It represents the downlink counterpart of the term $\gamma_{mk}$. 
The second term includes inter-user interference and uplink pilot contamination effect. The third term in the denominator is the variance of the normalized noise.

Although~\eqref{eq:DLrateCF} represents a further approximation, its importance is twofold: $(i)$ being in closed form, it allows the formulation of  power control policies that optimize the downlink rate for given  channel statistics. Hence, optimal power control can be performed on the large-scale fading time scale; $(ii)$ it gives relevant insights when compared to the expression of the achievable downlink rate in absence of downlink training, given in~\cite{Ngo2017b}.

\subsection{Use-and-Forget Achievable Downlink Rate}

The use-and-forget capacity-bounding technique~\cite{Marzetta2016a} allows us to obtain simpler, alternative  closed-form expression for the achievable rate.
By dividing the received data signal at UE $k$, given in~\eqref{eq:receiveddownlinksignal}, by $\sqrt{\Pd}\hat{a}_{kk}$, and adding and subtracting the term $ \EX{\frac{a_{kk}}{\hat{a}_{kk}}} q_{k}$,  we obtain
\begin{align}
\label{eq:normalized received signal}
r^\prime_{\mathrm{d},k} \triangleq \frac{r_{\mathrm{d},k}}{\sqrt{\Pd} \ \hat{a}_{kk}} &=\EX{\frac{a_{kk}}{\hat{a}_{kk}}}\! q_{k}\!+\!\left(\frac{a_{kk}}{\hat{a}_{kk}}\!-\! \EX{\frac{a_{kk}}{\hat{a}_{kk}}} \right)\! q_{k}\! \nonumber \\
&\quad+\! \sum\limits^K_{k' \neq k} \frac{a_{kk'}}{\hat{a}_{kk}}\! q_{k'}\! +\! \frac{\wdk}{\sqrt{\Pd} \hat{a}_{kk}}.
\end{align}  
Then, using the technique in \cite[Sec.~2.3.4]{Marzetta2016a} we obtain the following achievable downlink rate expression for the $k$th UE:
\begin{equation}
\label{eq:DLrateUnF} 
R_k^{\mathsf{UnF}} = \log_2 \!\left( \!1 \!+\!
\frac{\left|\EX{\frac{a_{kk}}{\hat{a}_{kk}}} \right|^2}{\varx{ \frac{a_{kk}}{\hat{a}_{kk}}} + \sum\limits^K_{k' \neq k} \EX{\left|\frac{a_{kk'}}{\hat{a}_{kk}}\right|^2} + \frac{1}{\Pd}\EX{ \frac{1}{|\hat{a}_{kk}|^2}}} \!\right).
\end{equation}
\begin{remark}
The use-and-forget bound gives a rigorous lower bound on capacity regardless of
the amount of channel hardening. However, it is tight only when there is significant
channel hardening~\cite{Marzetta2016a,Ngo2017a,Caire2017a}. (If the channel gain $\hat{a}_{kk}$ hardens then $\EX{a_{kk} \mathrel{\big|} \hat{a}_{kk}}$ is well approximated by $\EX{a_{kk}}$.)
Hence, the bound in~\eqref{eq:DLrateUnF} gives a pessimistic estimate of the achievable downlink rate when channel hardening does not occur, and  we expect the bound in~\eqref{eq:DLrateCF} to give us a larger achievable rate. 
\end{remark}

\subsection{Downlink Pilots, or No Downlink Pilots, that is the Question}
The achievable downlink rate assuming that UE $k$ has only access to the statistical CSI, hence denoted as $R_k^{\mathsf{sCSI}}$, is given by~\cite{Ngo2017b} and shown in~\eqref{eq:DLrateHien} at the top of the next page.
The first term in the denominator of the effective \SINR represents the so-called \textit{beamforming gain uncertainty} and comes from the users' lack of the CSI knowledge. The beamforming gain uncertainty, equal to the variance of the effective downlink channel, also gives an alternative measure of the channel hardening as the more the channel hardens, the smaller this term is.

We observe that~\eqref{eq:DLrateCF} and~\eqref{eq:DLrateHien} only differ by the term $\kappa_k$ which appears  both in the numerator and denominator of the effective \SINR but with different sign. The term $\kappa_k$, defined in~\eqref{eq:kappa-def}, conveys the contribution from the downlink training. It also
captures all the dependencies from the downlink pilot sequences, i.e., downlink pilot contamination terms. 
Being the variance of the downlink channel estimate, $\kappa_k$ is non-negative. 
Indeed, it can be also seen as a scaled version of the \textit{Pearson correlation coefficient} between $a_{kk}$ and $\check{y}_{\mathrm{dp},k}$: \addtocounter{equation}{1}
\begin{align}
\label{eq:corr-coefficient}
&\rho_{a_{kk},\check{y}_{\mathrm{dp},k}} \triangleq \frac{\cov{a_{kk},\check{y}_{\mathrm{dp},k}}}{\sqrt{\varx{a_{kk}} \varx{\check{y}_{\mathrm{dp},k}}}}
\nonumber \\ &\implies \kappa_k = \varx{a_{kk}}|\rho_{a_{kk},\check{y}_{\mathrm{dp},k}}|^2. 
\end{align}
By definition, $\rho_{a_{kk}\check{y}_{\mathrm{dp},k}}$ has a value between -1 and 1, thus $|\rho_{a_{kk}\check{y}_{\mathrm{dp},k}}|^2 \leq 1$, with equality if total positive linear correlation. It follows that $\kappa_k \leq \varx{a_{kk}}$, $\forall k$.

By using this inequality, a further upper bound on the achievable downlink rate is given by~\eqref{eq:DL-rate-closed-form} at the top of the next page.
Comparing expressions~\eqref{eq:DLrateCF} and~\eqref{eq:DLrateHien}, we note that the benefits introduced by the downlink training correspond to a coherent beamforming gain improved by an additive term $\kappa_k$, at most equal to the variance of the effective downlink channel, and a beamforming gain uncertainty replaced by a smaller term representing the variance of the downlink channel estimation error. 
\begin{remark}
If we assume mutually orthogonal downlink pilots, that is $\taudp = K$, then \addtocounter{equation}{1}
\begin{equation}
\kappa_k = \frac{\taudp \Pdp \varsigma_{kk}^2}{\taudp \Pdp \varsigma_{kk}+1} \rightarrow \varsigma_{kk}, \qquad \text{if } \taudp \Pdp \varsigma_{kk} \gg 1 .
\end{equation}  
\end{remark}
Generalizing, the variance of the downlink channel estimate, $\kappa_k$, tends to the variance of the effective downlink channel, $\varsigma_{kk}$, if the downlink training length or the power spent on downlink pilots is sufficiently large to overcome the downlink pilot contamination and to guarantee an excellent quality of the channel estimation. In this case,~\eqref{eq:DL-rate-closed-form} approaches~\eqref{eq:DLrateCF}. 

Importantly, if the channel seen by UE $k$ significantly hardens then $\varsigma_{kk} \rightarrow 0$ as well as $\kappa_k\to 0$, and all the capacity lower bounds presented in this section collapse to only one expression,~\eqref{eq:DLrateHien}, which would represent a very good estimate of the actual achievable downlink rate.

\begin{figure*}[!t]
\normalsize
\setcounter{eqcnt2}{\value{equation}}
\setcounter{equation}{34}
\begin{equation}
\label{eq:DLrateHien} R_k^{\mathsf{sCSI}} = \log_2 \!\left( \!1 \!+\!
\frac{\Pd\left( \sum\nolimits_{m=1}^M \sqrt{\eta_{mk}} \gamma_{mk} \right)^2}{\Pd\varsigma_{kk} + \Pd \sum\limits_{k' \neq k}^K \left[ \varsigma_{kk^\prime} + \left|\bm{\varphi}_{k'}\herm\bm{\varphi}_{k}\right|^2 \left(\sum\limits_{m=1}^M \sqrt{\eta_{mk'}} \gamma_{mk'} \frac{\beta_{mk}}{\beta_{mk'}} \right)^2 \right] + 1} \!\right).
\end{equation}
\setcounter{equation}{\value{eqcnt2}}
\end{figure*}

\begin{figure*}[!t]
\normalsize
\setcounter{eqcnt3}{\value{equation}}
\setcounter{equation}{36}
\begin{equation}
\label{eq:DL-rate-closed-form}
R_k^{\mathsf{cf}} \leq R_k^{\mathsf{ub}} \triangleq \log_2 \!\left( \!1 \!+\!
\frac{\Pd\left( \sum\nolimits_{m=1}^M \sqrt{\eta_{mk}} \gamma_{mk} \right)^2+ \Pd\varsigma_{kk}}{\Pd \sum\limits_{k' \neq k}^K \varsigma_{kk^\prime} + \Pd \sum\limits_{k' \neq k}^K \left|\bm{\varphi}_{k'}\herm\bm{\varphi}_{k}\right|^2 \left(\sum\limits_{m=1}^M \sqrt{\eta_{mk'}} \gamma_{mk'} \frac{\beta_{mk}}{\beta_{mk'}} \right)^2 + 1} \!\right).  
\end{equation}
\setcounter{equation}{\value{eqcnt3}}
\hrulefill
\vspace*{4pt}
\end{figure*}

\section{Resource Allocation}
\label{sec:resource-allocation}

\subsection{Max-Min Fairness Power Control}
\label{subsec:power-control}

Max-min fairness power control (MMF-PC) enables uniformly good service throughout the network. Specifically, the downlink power is allocated to maximize the minimum achievable downlink rate in the system.  
Such an egalitarian policy penalizes UEs with excellent channel condition, especially when  ``poor'' UEs are present. 
 
We rely on the closed-form expression~\eqref{eq:DLrateCF} to formulate the MMF-PC optimization problem.
Maximizing the minimum $R_k^{\mathsf{cf}}$ is equivalent to maximize the lowest $\SINR_k^\mathsf{cf}$ in the network. For the sake of brevity, let us define 
\[ \upsilon_{kk^\prime} \triangleq \left|\bm{\varphi}_{k'}\herm\bm{\varphi}_{k}\right|^2 \left(\sum\limits_{m=1}^M \sqrt{\eta_{mk'}} \gamma_{mk'} \frac{\beta_{mk}}{\beta_{mk'}} \right)^2. \]
The MMF-PC optimization problem, under per-AP power constraints, is given by
\begin{subequations} \label{Problem:Max-Min-Original-SINR}
\begin{align}	
  \mathop {\max}\limits_{\{\eta_{mk}\}} & \quad \min_{k} \frac{\Pd\left( \sum\nolimits_{m=1}^M \sqrt{\eta_{mk}} \gamma_{mk} \right)^2+ \Pd\kappa_k}{\Pd \sum\limits_{k'=1}^K \varsigma_{kk^\prime}-\Pd\kappa_k + \Pd \sum\limits_{k' \neq k}^K \upsilon_{kk^\prime} + 1} \label{eq:obj}\\
  \text{s.t.} &\quad \sum\nolimits_{k=1}^K \eta_{mk}\gamma_{mk} \leq 1, ~\forall m, \label{eq:maxpower1}\\
  &\quad \eta_{mk} \geq 0, ~ \forall k, ~ \forall m.\label{eq:power1}
  \end{align}
\end{subequations}
Problem~\eqref{Problem:Max-Min-Original-SINR} is nonconvex since~\eqref{eq:obj} is neither convex nor concave with respect to $\{\eta_{mk}\}$. The term $\kappa_k$ has a big impact on the tractability of the problem. If $\kappa_k = 0$,~\eqref{eq:obj} becomes the downlink \SINR expression without downlink training, and the corresponding optimization problem admits global optimal solutions that can be computed by a sequence of second-order cone programs (SOCPs)~\cite{Ngo2017b}. 
However, this optimization does not take into account the contribution from the downlink training. Our idea is to reformulate problem~\eqref{Problem:Max-Min-Original-SINR} as an SOCP by approximating~\eqref{eq:obj}, while preserving the downlink training gain. 
First, we use~\eqref{eq:DL-rate-closed-form} to approximate the problem as
\begin{subequations} \label{Problem:Max-Min-DL-rate-closed-form}
\begin{align}	
  \mathop {\max}\limits_{\{\eta_{mk}\}} & \quad \min_{k} \frac{\Pd\left( \sum\limits_{m=1}^M \sqrt{\eta_{mk}} \gamma_{mk} \right)^2+ \Pd\varsigma_{kk}}{\Pd \sum\limits_{k'=1}^K \varsigma_{kk^\prime} \!-\! \Pd \varsigma_{kk}\! +\! \Pd \sum\limits_{k' \neq k}^K \upsilon_{kk^\prime}\! +\! 1} \label{eq:obj-2}\\
  \text{s.t.} &\quad \eqref{eq:maxpower1}, \eqref{eq:power1}.
  \end{align}
\end{subequations}
Secondly, similar to~\cite{Ngo2017b}, we define $\zeta_{mk} \triangleq \sqrt{\eta_{mk}}$, and introduce slack variables $\vartheta_m$ and $\varrho_{k^\prime k}$ to reshape the problem as in~\eqref{Problem:Max-Min-Slack} at the top of the next page.  
Problems~\eqref{Problem:Max-Min-DL-rate-closed-form} and~\eqref{Problem:Max-Min-Slack} are equivalent as the first and second constraints hold with equality at the optimum.
The equivalent epigraph formulation of~\eqref{Problem:Max-Min-Slack} is  \addtocounter{equation}{1}
\begin{subequations}\label{Problem:Max-Min-Epigraph}
\begin{align} 
	\mathop {\max}\limits_{\{\zeta_{mk}, \varrho_{k^\prime k}, \vartheta_m\}, \nu} & \quad \nu \label{eq:obj-epigraph} \\	
	 \text{s.t.} &\quad \nu \cdot \norm{\bv_k}^2 \leq \left( \sum\nolimits^M_{m=1} \gamma_{mk} \zeta_{mk} \right)^2  \nonumber \\
	 &\qquad+ (1+\nu) \sum\nolimits_{m=1}^M \beta_{mk} \gamma_{mk} \zeta^2_{mk}, \; \forall k, \label{eq:constr-epigraph-b}\\ 
	&\quad \eqref{eq:constr-slack-b}, \eqref{eq:constr-slack-c}, \eqref{eq:constr-slack-d}, \eqref{eq:constr-slack-e}.
\end{align}
\end{subequations}
where $\bv_k \triangleq \left[ \bv_{k1}\trans \bI_{-k} \quad \bv_{k2}\trans \quad \frac{1}{\sqrt{\Pd}} \right]\trans$, $\bI_{-k} \in \C^{K \times (K-1)}$ is $\bI_K$ with $k$th column removed, $\bv_{k1} \triangleq \left[ \bm{\varphi}\herm_1 \bm{\varphi}_k \varrho_{1k} \ \cdots \ \bm{\varphi}\herm_K \bm{\varphi}_k \varrho_{Kk} \right]\trans$, and $\bv_{k2} \triangleq \left[ \sqrt{\beta_{1k}} \vartheta_1 \ \cdots \ \sqrt{\beta_{Mk}} \vartheta_M \right]\trans$.
The \SINR constraint~\eqref{eq:constr-epigraph-b}, is still neither convex nor concave with respect to $\zeta_{mk}$. 
To overcome such non-convexity, we use \textit{sequential convex approximation} (SCA).
Let us define the following vectors associated to UE $k$: $\bm{\gamma}_k =~[\gamma_{1k} \: \cdots \: \gamma_{Mk}]\trans$, $\bm{\bar{\gamma}}_k =~[\sqrt{\gamma_{1k}} \: \cdots \: \sqrt{\gamma_{Mk}}]\trans$, $\bm{\beta}_k =~[\beta_{1k} \: \cdots \: \beta_{Mk}]\trans$, $\bm{\bar{\beta}}_k =~[\sqrt{\beta_{1k}} \: \cdots \: \sqrt{\beta_{Mk}}]\trans$, and $\bm{\zeta}_k = ~[\zeta_{1k} \; \cdots \; \zeta_{Mk}]\trans$. We rewrite the right-hand side of~\eqref{eq:constr-epigraph-b} as $f(\bm{\zeta}_k) \triangleq \left( \bm{\gamma}\trans_k \bm{\zeta}_k \right)^2 + (1+\nu) \norm{\bm{\bar{\gamma}}_k \circ \bm{\bar{\beta}}_k \circ \bm{\zeta}_k}^2$.
We form a convex approximation $\hat{f}$ of $f$ by using a first-order Taylor expansion. Let $\zeta^n_{mk}$ be the value of $\zeta_{mk}$ at the $n$th iteration of the SCA algorithm, and $\bm{\zeta}^n_k = [\zeta^n_{1k} \; \cdots \; \zeta^n_{Mk}]\trans$ be the corresponding vector associated to UE $k$, $\hat{f}$ is given by  
\begin{align*}
\hat{f}(\bm{\zeta}_k;\bm{\zeta}^n_k) &= f(\bm{\bm{\zeta}^n_k}) + (\bm{\zeta}_k-\bm{\zeta}^n_k)\trans \nabla f(\bm{\zeta}^n_k) \nonumber \\
&= (1+\nu)\norm{\bm{\bar{\gamma}}_k \circ \bm{\bar{\beta}}_k \circ \bm{\zeta}^n_k}^2 + 2 \bm{\zeta}\trans_k \bm{\gamma}_k  \bm{\gamma}\trans_k \bm{\zeta}^n_k - \norm{\bm{\gamma}\trans_k \bm{\zeta}^n_k}^2 \nonumber \\
&\quad+ 2 (1+\nu) (\bm{\zeta}_k - \bm{\zeta}^n_k)\trans (\bm{\gamma}_k \circ \bm{\beta}_k \circ \bm{\zeta}^n_k).
\end{align*}
The \SINR constraint~\eqref{eq:constr-epigraph-b} can be rewritten as
\[\nu \cdot \norm{\bv_k}^2 \leq \hat{f}(\bm{\zeta}_k;\bm{\zeta}^n_k)\] and, equivalently, as
\begin{align}
\label{eq:hyperbolic-constraint}
\norm{\bar{\bv}_k}^2 \! &\leq \! \left(1+\frac{1}{\nu}\right) \! \left( \norm{\bm{\bar{\gamma}}_k \! \circ \! \bm{\bar{\beta}}_k \! \circ \! \bm{\zeta}^n_k}^2 \! + \! \frac{2}{1+\nu} \bm{\zeta}\trans_k \bm{\gamma}_k  \bm{\gamma}\trans_k \bm{\zeta}^n_k \! \right. \nonumber \\
&\left. \quad +  2 (\bm{\zeta}_k \! - \! \bm{\zeta}^n_k)\trans (\bm{\gamma}_k \! \circ \! \bm{\beta}_k \! \circ \! \bm{\zeta}^n_k) \right),
\end{align}
where we defined $\bar{\bv}_k \triangleq \left[\bv_{k1}\trans \bI_{-k} \quad \bv_{k2}\trans \quad \frac{1}{\sqrt{\nu}}\bm{\gamma}_k\trans \bm{\zeta}^n_k \quad \frac{1}{\sqrt{p_d}} \right]\trans$.

\begin{figure*}[!t]
\normalsize
\setcounter{eqcnt4}{\value{equation}}
\setcounter{equation}{40}
\begin{subequations}\label{Problem:Max-Min-Slack}
\begin{align} 
	\mathop {\max}\limits_{\{\zeta_{mk}, \varrho_{k^\prime k}, \vartheta_m\}} & \quad \min_{k} \frac{\left( \sum\nolimits_{m=1}^M \gamma_{mk} \zeta_{mk} \right)^2 + \sum\nolimits_{m=1}^M \beta_{mk} \gamma_{mk} \zeta^2_{mk}}{ \sum\limits_{m=1}^M  \beta_{mk} \vartheta^2_m - \sum\limits_{m=1}^M \beta_{mk} \gamma_{mk} \zeta^2_{mk}  + \sum\limits_{k^\prime \neq k}^K \left|\bm{\varphi}_{k'}\herm\bm{\varphi}_{k}\right|^2 \varrho^2_{k^\prime k} + \frac{1}{\Pd}} \label{eq:obj-slack} \\	
	 \text{s.t.} &\quad \sum\nolimits_{k^\prime = 1}^K \gamma_{mk^\prime} \zeta^2_{mk^\prime} \leq \vartheta^2_m, \; \forall m, \label{eq:constr-slack-b}\\ 
	&\quad \sum\nolimits_{m=1}^M \gamma_{mk'} \frac{\beta_{mk}}{\beta_{mk'}} \zeta_{mk'}  \leq \varrho_{k^\prime k}, \; \forall k^\prime \neq k, \label{eq:constr-slack-c}\\
	&\quad 0 \leq \vartheta_m \leq 1, \; \forall m,\label{eq:constr-slack-d}\\
	&\quad \zeta_{mk} \geq 0, \; \forall m, \; \forall k \label{eq:constr-slack-e}.
\end{align}
\end{subequations}
\setcounter{equation}{\value{eqcnt4}}
\hrulefill
\vspace*{4pt}
\end{figure*}

Expression~\eqref{eq:hyperbolic-constraint} describes a \textit{hyperbolic constraint} and represents a class of convex problems that can be cast as SOCPs~\cite{Lobo1998}, by using
\begin{equation}
\label{eq:hyperbolic-constraint-to-SOCP}
\begin{aligned}
	\norm{\bA \bx\!+\!\bb}^2 & \leq t(\bc\trans\bx \! + \! d)\\
	\bc\trans \bx + d & \geq 0 \\
	t & \geq 0
\end{aligned} \!\iff \!
\norm{\begin{bmatrix}
	2(\bA \bx\!+\!\bb) \\
	\bc\trans \bx\! + d \!- \!t \\
\end{bmatrix}} \leq \bc\trans \bx\! +\! d \!+\! t. 
\end{equation}
Letting $\bx = \bm{\zeta}_k$, by comparing~\eqref{eq:hyperbolic-constraint} and~\eqref{eq:hyperbolic-constraint-to-SOCP}, we obtain that $\bA \bx+\bb = \bar{\bv}_k$, $t = 1 + \frac{1}{\nu}$, and 
$\bc\trans \bx \!+\! d \triangleq s_k = \frac{2}{1+\nu}~(\bm{\zeta}^n_k)\trans \bm{\gamma}_k  \bm{\gamma}\trans_k \bm{\zeta}_k\!+\! \norm{\bm{\bar{\gamma}}_k \! \circ \! \bm{\bar{\beta}}_k \! \circ \! \bm{\zeta}^n_k}^2 \! + 2 (\bm{\zeta}_k - \bm{\zeta}^n_k)\trans (\bm{\gamma}_k  \circ  \bm{\beta}_k  \circ  \bm{\zeta}^n_k)$.
This gives the following optimization problem
\begin{subequations}\label{Problem:Max-Min-Final}
\begin{align} 
	&\mathop {\max}\limits_{\{\zeta_{mk}, \varrho_{k^\prime k}, \vartheta_m\}, \nu} \quad \nu \label{eq:obj-final} \\	
	&\qquad \text{ s.t.} \quad \norm{\begin{bmatrix}
	2\bv_{k1}\trans \bI_{-k} \quad 2\bv_{k2}\trans \quad \frac{2}{\sqrt{\nu}}\bm{\gamma}_k\trans \bm{\zeta}^n_k \quad \frac{2}{\sqrt{p_d}} \quad s_k\!-\!\left(1\!+\!\frac{1}{\nu}\right)
	\end{bmatrix}\trans} \nonumber \\
	& \qquad\qquad\qquad \leq \! s_k \!+ \! 1 \!+ \!1/\nu, \; \forall k, \label{eq:constr-final-b}\\
	&\qquad \qquad \ s_k \geq 0, \; \forall k, \label{eq:constr-final-c}\\  
	&\qquad \qquad \ \eqref{eq:constr-slack-b}, \eqref{eq:constr-slack-c}, \eqref{eq:constr-slack-d}, \eqref{eq:constr-slack-e}. \label{eq:constr-final-d}
\end{align}
\end{subequations}

If $\nu$ is fixed, problem~\eqref{Problem:Max-Min-Final} is a convex program, and the optimal solution can be efficiently computed by using interior-point methods,
for example with the toolbox CVX~\cite{cvx2015}. Letting $\nu$ vary over an \SINR search range $\{\nu_{\mathsf{min}},\nu_{\mathsf{max}}\}$, the optimal solution can be efficiently computed by using the \textit{bisection method}~\cite{Boyd2004a}, in each step solving the following convex feasibility problem, for a fixed value of $\nu$,
\begin{equation}
\label{Problem:feasibility-problem}  
\begin{aligned}
	\mathop \text{find} & \quad \{\bm{\zeta}_k\}, \{\bm{\varrho}_k\}, \{\vartheta_m\} \\	
	 \text{s.t.} & \quad \eqref{eq:constr-final-b}, \eqref{eq:constr-final-c}, \eqref{eq:constr-final-d}.
\end{aligned}
\end{equation}
where $\bm{\varrho}_k = [\varrho_{1k} \: \cdots \: \varrho_{Kk}] \bI_{-k}$.
The SCA algorithm to solve~\eqref{Problem:Max-Min-Epigraph} is described in Algorithm~\ref{alg:SCA}.

\begin{algorithm}[!t]
\small
\setstretch{1}
\caption{SCA algorithm for Max-Min Power Control}
\vspace{1mm}
\textbf{Result:} Solve optimization in~{\eqref{Problem:Max-Min-DL-rate-closed-form}}. \\
\textbf{Input:} Initial power control coefficient vector $\bm{\zeta}^0_k$, initial \SINR upper bound $\nu_{\mathsf{max}}$, $\nu_{\mathsf{min}}=0$, line-search accuracy $\epsilon$, maximum number of iteration $N_{\text{I}}$, $n=0$.
\begin{algorithmic}[1]
\While {$n<N_{\text{I}}$}
\While {$\nu_{\mathsf{max}}-\nu_{\mathsf{min}}>\epsilon$}
\State Set $\nu = (\nu_{\mathsf{max}} + \nu_{\mathsf{min}})/2$;
\State Solve~\eqref{Problem:feasibility-problem}; \textbf{Input:} $\bm{\zeta}^n_k$.
\If {\eqref{Problem:feasibility-problem} is feasible for $\nu, \forall k$}
\State Set $\nu_{\mathsf{min}}=\nu$; Set $\bm{\zeta}^\ast_k$ as the solution to~\eqref{Problem:feasibility-problem}.
\Else
\State Set $\nu_{\mathsf{max}}=\nu$;
\EndIf
\EndWhile
\State Set $n=n+1$; Set $\bm{\zeta}^n_k = \bm{\zeta}^\ast_k$;
\EndWhile
\end{algorithmic}
\textbf{Output:} $\bm{\zeta}^n_k$.
\label{alg:SCA}
\end{algorithm}

The optimal power control coefficients, output of Algorithm~\ref{alg:SCA}, are then inserted into~\eqref{eq:DLrateCF} to evaluate the actual system performance. Clearly, these coefficients are sub-optimal for~\eqref{Problem:Max-Min-Original-SINR}.

\subsection{Pilot Sequence Assignment}
\label{subsec:pilot-assignment}

The CPU is responsible, among others, to establish the uplink/downlink training time length, to assign the pilot sequences to the UEs and inform the APs about this mapping. 
When pilot reuse is adopted, pilot assignment strategies become relevant to mitigate the pilot contamination and improve system performance.

Unlike the simple random pilot assignment method, advanced pilot assignment policies usually require CSI to estimate and optimize a QoS-based objective function. To reduce this increase of signaling overhead over the fronthaul network, pilot assignment can be designed to depend only on the long-term channel statistics. Hence, downlink power control and pilot assignment can be performed jointly on a slow time scale, i.e., over multiple coherence intervals. Importantly, they can be also combined since the downlink pilot transmission is power controlled. 

Before investigating advanced pilot assignment strategies, we first describe a method that satisfies the constraint in~\eqref{eq:pilot-assignment-constraint}, referred to as \textit{baseline joint uplink (UL) and downlink (DL) pilot assignment}. 
The CPU assigns mutually orthogonal downlink pilots to the UEs using the same uplink pilots.
Let $\mathcal{U} = \{ 1, \ldots, \taup \}$ and $\mathcal{D} = \{ 1, \ldots, \taudp \}$ be the set of indices of the uplink and downlink pilots, respectively. 
The set of all ordered pairs of uplink and downlink pilot indices is given by the Cartesian product $\mathcal{U} \times \mathcal{D}$. 
Each pair represents a possible choice of uplink and downlink pilots that satisfies~\eqref{eq:pilot-assignment-constraint}. 
The number of available pairs is equal to $|\mathcal{U} \times \mathcal{D}| = |\mathcal{U}| \cdot |\mathcal{D}| =~\taup \taudp$. 
Hence, the training lengths must be set to ensure that $\taup \taudp\geq~K$. 
Notice that, $\taup, \taudp$ are upper bounded by the length of the coherence interval $\tauc$, thus if $K$ is too large (compared to $\tauc$), setting $\taup, \taudp$ accordingly might not be efficient, or even possible.
However, in practical cases of interest $K \ll \tauc$.
Importantly, the set $\mathcal{U} \times \mathcal{D}$ can be computed off-line by the CPU, once $\taup$ and $\taudp$ are determined. 
The baseline joint uplink and downlink pilot assignment strategy, described in Algorithm~\ref{alg:baseline-pilot-assignment}, consists in randomly assigning a pair of uplink and downlink pilots from the product set $\mathcal{U} \times \mathcal{D}$ to all the UEs, assuming that $\taup \taudp \geq K$.
If $K > \taup \taudp$, then $K-\taup \taudp$ UEs are scheduled in the next coherent interval.
\begin{algorithm}[!t]
\small
\setstretch{1}
\caption{Baseline joint UL and DL pilot assignment}
\vspace{1mm}
\textbf{Result:} UL and DL pilot sequence assignment for all the users \\
\textbf{Input:} $\mathcal{U}, \mathcal{D}, K$, UL pilot book $\mathcal{P}^u$, DL pilot book $\mathcal{P}^d$.
\begin{algorithmic}[1]
\If {$|\mathcal{U}| \cdot |\mathcal{D}| \geq K$}
\State $\mathcal{T} = \mathcal{U} \times \mathcal{D}$;
\For {$k=1:K$}
\State Set $r \in \mathbb{Z} \cap [1, |\mathcal{T}|]$;  \; // $r$ is drawn at random
\Statex // the $r$-th element of $\mathcal{T}$ is a pair of pilot indices
\State Set $(i, j) = [\mathcal{T}]_r$; 
\State Set $(\bm{\varphi}_k, \bm{\psi}_k) = ([\mathcal{P}^u]_i,[\mathcal{P}^d]_j)$;
\State Set $[\mathcal{T}]_r = [\quad]$; \; // remove element $r$ from $\mathcal{T}$
\EndFor
\EndIf
\end{algorithmic}
\textbf{Output:} $\{\bm{\varphi}_k$,$\bm{\psi}_k\},~\forall k$.
\label{alg:baseline-pilot-assignment}
\end{algorithm}

Finding the optimal pair of uplink and downlink pilot sequences is a difficult combinatorial problem. Here we only focus on \textit{greedy} algorithms that simplify the computation while introducing a significant performance gain. We propose two pilot assignment methods:
\begin{itemize}
\item \textit{Greedy} (Algorithm~\ref{alg:greedy}): Uplink and downlink pilots are initially assigned by performing Algorithm~\ref{alg:baseline-pilot-assignment}. Then, they are jointly selected to minimize the pilot contamination of the UE with the lowest downlink rate. More specifically, given the UE $\hat{k}$ with lowest rate, the utility function corresponds to the sum between the mean-square of the downlink pilot contamination term in~\eqref{eq:y_k}, and the mean-square of the uplink pilot contamination term in~\eqref{eq:uplinkpilotprojection}, summed over all APs. 
A similar method has been proposed in~\cite{Ngo2017b} for the uplink pilot selection. Here, it is extended to the downlink pilot selection as well. Moreover, the choice of the uplink and downlink pilot pairs is constrained to be within the set $\mathcal{U} \times \mathcal{D}$. If the optimal pair of pilots for the worst UE had been assigned to another UE at a first stage, then the CPU swaps the pilot assignments of those two UEs. The final if/else statement in Algorithm~\ref{alg:greedy} ensures convergence: the final minimum rate produced by Algorithm~\ref{alg:greedy} is larger than or equal to the initial minimum rate produced by Algorithm~\ref{alg:baseline-pilot-assignment}.
\item \textit{Advanced Greedy} (Algorithm~\ref{alg:advanced-greedy}): It only differs from the greedy approach for the utility function set to the highest minimum downlink rate among the UEs. Convergence of Algorithm~\ref{alg:advanced-greedy} is ensured by the fact that at each iteration $n$, the pilot pair of the worst-rate UE is selected to maximize the minimum rate among all the UEs (after swapping).
\end{itemize}
\begin{algorithm}[!t]
\small
\setstretch{1.1}
\caption{Greedy algorithm for UL and DL pilot assignment}
\vspace{1mm}
\textbf{Result:} UL and DL pilot sequence assignment for all the users\\ \hfill
\textbf{Input:} $\mathcal{T}= \mathcal{U} \times \mathcal{D}$, UL pilot book $\mathcal{P}^u$, DL pilot book $\mathcal{P}^d$, $\{\bm{\varphi}^0_k$,$\bm{\psi}^0_k\}$ as output of Algorithm~\ref{alg:baseline-pilot-assignment}. Number of iterations $N_{\text{I}}$, $n=0$. \\ 
Utility function:\\$f(k,\!\{\bm{\varphi}_k,\!\bm{\psi}_k\})\!=\!\sum\limits_{m=1}^M \sum\limits_{k' \neq k}^K \left( \beta_{mk'} \left|\bm{\varphi}_{k}\herm\bm{\varphi}_{k'}\right|^2\!+\!\eta_{mk'} \beta_{mk} \gamma_{mk'}\left|\bm{\psi}_k\herm \bm{\psi}_{k'}\right|^2\right)$.
\begin{algorithmic}[1]
\While {$n < N_{\text{I}}$}
\State Set $n = n+1$;\; Set $\{\bm{\varphi}^n_k, \bm{\psi}^n_k\} = \{\bm{\varphi}^{n-1}_k, \bm{\psi}^{n-1}_k\}$;
\State Set $\hat{k}^n = \underset{k}{\argmin}~R^{\mathsf{cf}}_k(\{\bm{\varphi}^n_k, \bm{\psi}^n_k\})$;
\For {$t=1:|\mathcal{T}|$}
\State Set $\{\bm{\tilde{\varphi}}^n_k, \bm{\tilde{\psi}}^n_k\} = \{\bm{\varphi}^n_k, \bm{\psi}^n_k\}$ // temporary assignment
\Statex // the $t$-th element of $\mathcal{T}$ is a unique pair of pilot indices
\State Set $(i, j) = [\mathcal{T}]_t$;
\Statex // find the UE that uses the pilot pair $(i,j)$
\State Set $k^*=\underset{k}{\text{find}} \left( (\bm{\tilde{\varphi}}^n_k, \bm{\tilde{\psi}}^n_k) == ([\mathcal{P}^u]_i,[\mathcal{P}^d]_j) \right)$;
\Statex // swap the pilot assignments if the pair $(i,j)$ is used
\IfThen{$k^* \neq \emptyset$}{Set $(\bm{\tilde{\varphi}}^n_{k^*}, \bm{\tilde{\psi}}^n_{k^*}) = (\bm{\tilde{\varphi}}^n_{\hat{k}^n}, \bm{\tilde{\psi}}^n_{\hat{k}^n})$;}
\State Set $(\bm{\tilde{\varphi}}^n_{\hat{k}^n}, \bm{\tilde{\psi}}^n_{\hat{k}^n}) = ([\mathcal{P}^u]_i,[\mathcal{P}^d]_j)$;
\State Set $\hat{f}_{i,j} = f(\hat{k}^n,\{\bm{\tilde{\varphi}}^n_k,\bm{\tilde{\psi}}^n_k\})$;
\EndFor
\State Set $(i^*, j^*) = \underset{i,j}{\argmin}~\hat{f}_{i,j}$;
\State Set $k^*=\underset{k}{\text{find}} \left( (\bm{\varphi}^n_k, \bm{\psi}^n_k) == ([\mathcal{P}^u]_{i^*},[\mathcal{P}^d]_{j^*}) \right)$;
\IfThen{$k^* \neq \emptyset$}{Set $(\bm{\varphi}^n_{k^*}, \bm{\psi}^n_{k^*}) = (\bm{\varphi}^n_{\hat{k}^n}, \bm{\psi}^n_{\hat{k}^n})$};
\State Set $(\bm{\varphi}^n_{\hat{k}^n}, \bm{\psi}^n_{\hat{k}^n}) = ([\mathcal{P}^u]_{i^*},[\mathcal{P}^d]_{j^*})$;
\EndWhile
\end{algorithmic}
$\begin{aligned}
\textbf{Output:}
\begin{cases}
\{\bm{\varphi}^n_k,\bm{\psi}^n_k\},\; \text{if } \min\limits_k R^{\mathsf{cf}}_k(\{\bm{\varphi}^n_k, \bm{\psi}^n_k\}) \geq \min\limits_k R^{\mathsf{cf}}_k(\{\bm{\varphi}^0_k, \bm{\psi}^0_k\}), \\
\{\bm{\varphi}^0_k,\bm{\psi}^0_k\},\; \text{otherwise}.
\end{cases}
\end{aligned}$
\label{alg:greedy}
\end{algorithm}
\begin{algorithm}[!t]
\small
\setstretch{1}
\caption{Advanced greedy algorithm for UL and DL pilot assignment}
\vspace{1mm}
\textbf{Result:} UL and DL pilot sequence assignment for all the users \\
\textbf{Input:} $\mathcal{T}= \mathcal{U} \times \mathcal{D}$, UL pilot book $\mathcal{P}^u$, DL pilot book $\mathcal{P}^d$, $\{\bm{\varphi}^0_k$,$\bm{\psi}^0_k\}$ as output of Algorithm~\ref{alg:baseline-pilot-assignment}. Number of iterations $N_{\text{I}}$, $n=0$.\\ Utility function: $f(k,\{\bm{\varphi}_k,\bm{\psi}_k\}) = - \min\limits_k R_k^{\mathsf{cf}}(\{\bm{\varphi}_k, \bm{\psi}_k\})$.
\begin{algorithmic}[1]
\State \textbf{call} Algorithm~\ref{alg:greedy}
\end{algorithmic}
\vspace*{1mm}
\textbf{Output:} $\{\bm{\varphi}^n_k$,$\bm{\psi}^n_k\}$.
\label{alg:advanced-greedy}
\end{algorithm}

\section{Numerical Results and Discussions} \label{sec:num-results}
In this section, we numerically evaluate the performance of cell-free Massive MIMO with downlink beamforming training, in terms of \textit{per-user downlink net rate} (bits/s/Hz), defined as
\begin{equation} \label{eq:net-rate}
\bar{R}_k = \xi^{\mathsf{DL}}\left(1-\frac{\tau_{\mathrm{o}}}{\tauc}\right)R_k,
\end{equation}
where $\xi^{\mathsf{DL}}$ is the fraction of data symbols that are used for downlink payload transmission. $\tau_{\mathrm{o}}$ is the pilot overhead: $\tau_{\mathrm{o}}=\taup + \taud$ if downlink training is performed; $\tau_{\mathrm{o}}=\taup$ otherwise. 

\subsection{Simulation Scenario} \label{subsec:sim-scenario}
We assume that the $M$ APs and the $K$ UEs are uniformly distributed at random, within a square of size $D \times D$ km$^2$. A wrap-around technique is then implemented in order to simulate a cell-free network, eliminating border effects.
The large-scale fading consists in three slope path-loss model and log-normal shadow fading. For the sake of brevity, we do not describe this model in detail, but we refer to~\cite{Ngo2017b}. We also adopt the same simulation settings as in~\cite{Ngo2017b}. Unless otherwise stated, we consider: $D=1$ km; carrier frequency 2 GHz; $\tauc = 200$; $\xi^{\mathrm{DL}} = 0.5$; bandwidth 20 MHz; noise figure 9 dB; maximum radiated power 200 mW and 100 mW for downlink and uplink, respectively. In all examples, we take $K \ll \tau_c$ and $\taup\taudp \geq K$. Hence, Algorithm~\ref{alg:baseline-pilot-assignment} can be performed, and the decrease of freedom when assigning the pilot sequences, imposed by~\eqref{eq:pilot-assignment-constraint}, does not represent a serious issue.

\subsection{Pilot Allocation}
\begin{figure}[!t] \centering
	\subfloat[Per-user average downlink rate against training duration.\label{fig:fig1a}]{ 
    	\includegraphics[width=.9\columnwidth]{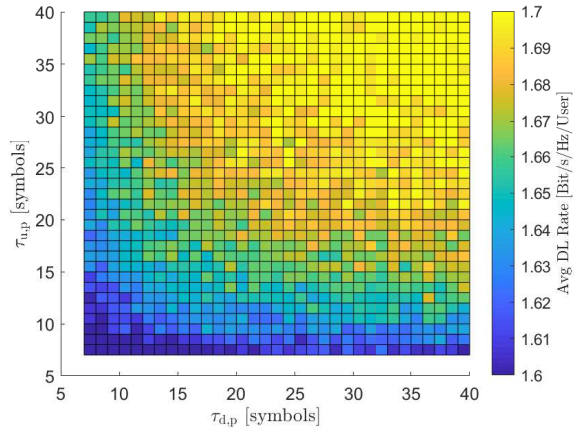}}\hfill
    \subfloat[Per-user average downlink net rate against training duration.\label{fig:fig1b}]{
    	\includegraphics[width=.9\columnwidth]{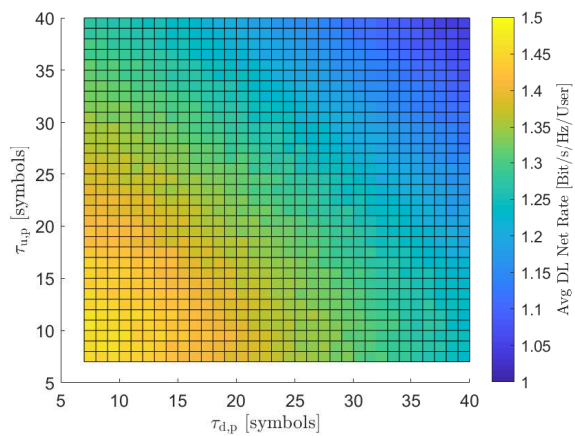}}\hfill
	\caption{The achievable downlink rate is affected equally by the quality of the uplink and downlink channel estimation, which is proportional to the length of the pilot sequences. However, pilot overhead reduces the net rate. Here, $M=200$, $K=40$.}
	\label{fig:fig1}  
\end{figure}

The uplink and downlink training duration can be conveniently adapted in order to maximize the system performance. For instance, the CPU can perform a search over all possible pilot lengths to maximize a utility such as max-min rate, average rate or sum rate. This method is optimal but the complexity grows exponentially with the number of UEs. \Figref{fig:fig1} shows the achievable downlink (gross and net) rate for different uplink and downlink pilot sequence lengths, assuming $M=200$, $K = 40$. 
Pilot sequences are assigned according to Algorithm~\ref{alg:baseline-pilot-assignment}, and it holds $\taup\taudp \geq K$. 
The rates are averaged over several random realizations of APs, UEs and shadowing. In these simulations, we assume that the CPU estimates the rates by using the closed-form expression given in~\eqref{eq:DLrateCF}. Furthermore, the power control coefficients at the $m$th AP are set as $\eta_{mk}=\left(\sum_{k^\prime=1}^K \gamma_{mk^\prime}\right)^{-1}~\forall k$, to satisfy~\eqref{eq:pwConstraintGamma} with equality. Clearly, this policy, herein referred to as \textit{channel-dependent full power transmission} (CD-FPT), is not optimal but it can be performed distributedly, and it speeds up the computation, since no optimization problem is involved.

In Fig.~\subref*{fig:fig1a}, we analyze the per-user average (gross) downlink rate, defined as $\frac{1}{K}\xi^{\mathsf{DL}}\sum_{k=1}^K R_k^{\mathsf{cf}}$. Clearly, the quality of the channel estimation is directly proportional to the training duration. In addition, the longer the pilot sequence is, the larger the number of mutually orthogonal pilots we can afford, reducing the pilot contamination. However, pilot overhead has a big impact on the actual rate provided to the UEs, especially when the ratio $\tauc/\tau_{\mathrm{o}}$ is small, as shown in Fig.~\subref*{fig:fig1b}. Indeed, we observe that the downlink net rate, defined as $\frac{1}{K}\sum_{k=1}^K \bar{R}_k^{\mathsf{cf}}$, is larger when $\taup$, $\taudp$ are in the range 7--20. Importantly, we note that $\taup$ plays a slightly more important role than $\taudp$. The uplink channel estimates are employed to define the precoders, leveraging the channel reciprocity. Hence, the channel estimation error and the uplink pilot contamination also propagate in the downlink, affecting the downlink channel estimates and the downlink pilot contamination. In fact, $\kappa_k$ depends on $\gamma_{mk}$ which in turn depends on the correlation between uplink pilots. 

\begin{figure}[!t] \centering
	\includegraphics[width=.9\columnwidth]{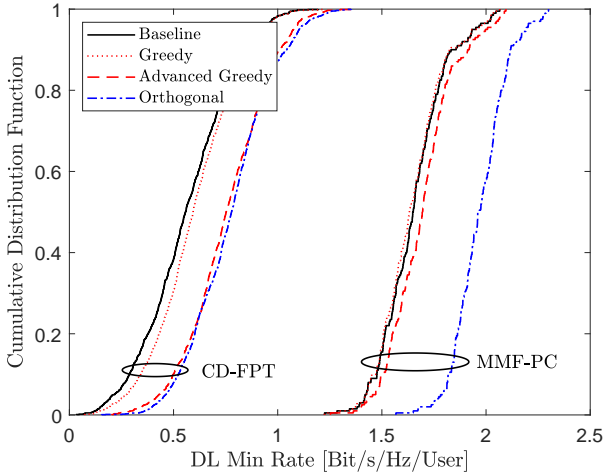} \vspace{-2mm}
	\caption{Per-user downlink minimum rate, with MMF-PC and CD-FPT, for different pilot assignment methods. Here, $N_{\mathrm{I}}\!=\!5$, $M\!=\!100$, $K\!=\!20$, and $\taup\!=\!\taudp\!=\!10$. For the orthogonal pilot assignment method $\taup\!=\!\taudp\!=\!20$.}  
	\label{fig:fig3}
\end{figure}

Once the optimal pilot lengths are selected, the CPU needs to properly assign the uplink and downlink pilot sequences to the UEs. \Figref{fig:fig3} compares the cumulative distribution functions (cdfs) of the downlink (gross) minimum rate obtained from the pilot assignment methods described in~\Secref{sec:resource-allocation}. The performance of both the MMF-PC, described in Algorithm~\ref{alg:SCA}, and the CD-FPT are evaluated.
 
Obviously, the baseline pilot assignment does not mitigate the interference from pilot contamination.
With CD-FPT, the greedy approach can provide up to 30\% gain, while the advanced greedy method approaches the upper bound, namely the orthogonal pilot assignment method (each UE has an orthogonal unique pilot sequence). 
In such a scenario, it is more likely that a poor rate is due to path-loss attenuation rather than pilot contamination. Even assigning the optimal pilot sequence pair that minimizes the pilot contamination might not be sufficient to significantly increase the lowest rate in the network.
With MMF-PC, the gap between the greedy methods and the baseline approach reduces.
MMF-PC inherently helps to alleviate the pilot contamination by setting different power levels for data and pilot transmission in the downlink to maximize the minimum rate. Hence, the greedy pilot assignment has less impact on the performance.   
The advanced greedy pilot assignment can still provide up to 4\% improvement of the minimum rate, as it aims to directly maximizes the lowest rate. The orthogonal pilot assignment method is, in this case, significantly better but it introduces additional pilot overhead and performs worse in terms of net downlink rate (see~\Figref{fig:fig5}).  

\subsection{Performance of the SCA Algorithm for Max-Min Power Control}

Let $\{\eta_{mk}^{\mathsf{sCSI}}\}$ be the power control coefficients that are solution of the optimization problem formulated in~\cite{Ngo2017b}, and $\{\eta_{mk}^{\mathsf{SCA}}\}$ the output of the SCA algorithm, described in Algorithm~\ref{alg:SCA}, which solves problem~\eqref{Problem:Max-Min-DL-rate-closed-form}. We insert these two sets of coefficients into~\eqref{Problem:Max-Min-Original-SINR} and look at the average minimum rate. 
\begin{figure}[!t] \centering
	\includegraphics[width=.9\columnwidth]{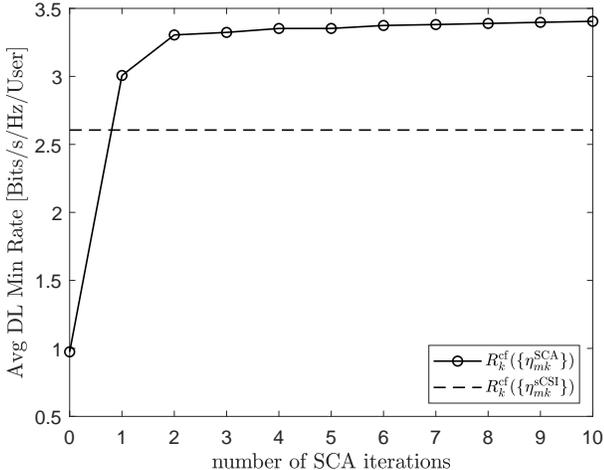} \vspace{-2mm}
	\caption{Per-user average downlink minimum rate, after MMF-PC, against number of SCA iterations. $\{\eta_{mk}^{\mathsf{sCSI}}\}$, $\{\eta_{mk}^{\mathsf{SCA}}\}$ are the power control coefficients output of the optimization problem defined in~\cite{Ngo2017b} and in Algorithm~\ref{alg:SCA}, respectively. Algorithm~\ref{alg:SCA} converges very quickly, providing much higher minimum rate. Here, $M=100$, $K=20$, and $\taup = \taudp = 10$.}  
	\label{fig:fig2}
\end{figure}
From~\Figref{fig:fig2}, we assert two remarkable facts:
\begin{itemize}
\item The SCA algorithm converges very quickly to the optimum. Very few SCA iterations are enough to solve the MMF-PC optimization problem and guarantee uniform service throughout the network. The power control coefficients at the $m$th AP, in the SCA iteration 0 are set as $\eta_{mk}^{\mathsf{SCA}}=\left(\sum_{k^\prime=1}^K \gamma_{mk^\prime}\right)^{-1}~\forall k$.   
\item The average downlink minimum rate obtained by using $\{\eta_{mk}^{\mathsf{SCA}}\}$ is significantly larger than the rate achieved by using the power control coefficients $\{\eta_{mk}^{\mathsf{sCSI}}\}$. This is due to the presence of the downlink training contribution $\kappa_k$ in the optimization problem~\eqref{Problem:Max-Min-Original-SINR}. Conversely, $\{\eta_{mk}^{\mathsf{sCSI}}\}$ are obtained by setting $\kappa_k = 0~\forall k$, and solving problem~\eqref{Problem:Max-Min-Original-SINR} in one iteration~\cite{Ngo2017b}.
\end{itemize}

\subsection{Downlink Training Gain}
In this section, we evaluate the achievable downlink rates presented in~\Secref{sec:performance-analysis}. We also compare the approximate achievable downlink rate proposed in~\eqref{eq:DLrateCF} with an alternative capacity lower bound that assumes no instantaneous CSI at the UE \cite[Lemma 3]{Caire2017a}. This bound is 
\begin{align} \label{eq:caire-lower-bound}
R_k^{\mathsf{lb}} &= \EX{\log_2 \left( 1 + \frac{\Pd |a_{kk}|^2}{\Pd \sum\nolimits_{k^\prime \neq k}^K |a_{kk^\prime}|^2+1} \right)} \nonumber \\
&\qquad- \frac{1}{\taud} \sum\limits_{k^\prime = 1}^K \log_2 \left( 1 + \taud \Pd \varx{a_{kk^\prime}} \right),
\end{align}    
which is essentially the achievable downlink rate for the case of perfect CSI knowledge at the UE (first term) decreased by a term that accounts for the lack of instantaneous CSI. The latter decreases as the coherence interval length  $\taud$ grows. This bound
can be  significantly tighter than~\eqref{eq:DLrateHien} especially when $\taud \gg K$ and 
there is lack of hardening~\cite{Caire2017a}. 
We next evaluate the achievable downlink net
 rates at an operational point where $ R_k^{\mathsf{lb}}$ is meaningful, that is, $\taud$ is large. We took  $\xi^{\mathsf{DL}}=1$, $\tauc = 500$ and $M \geq 100$. To ensure a fair comparison, we set the power control coefficients for all the cases as $\eta_{mk}=~\left(\sum_{k^\prime=1}^K \gamma_{mk^\prime}\right)^{-1}~\forall m,~\forall k$ (CD-FPT).  

\begin{figure}[!t] \centering
	\vspace*{-3mm}
	\includegraphics[width=.9\columnwidth]{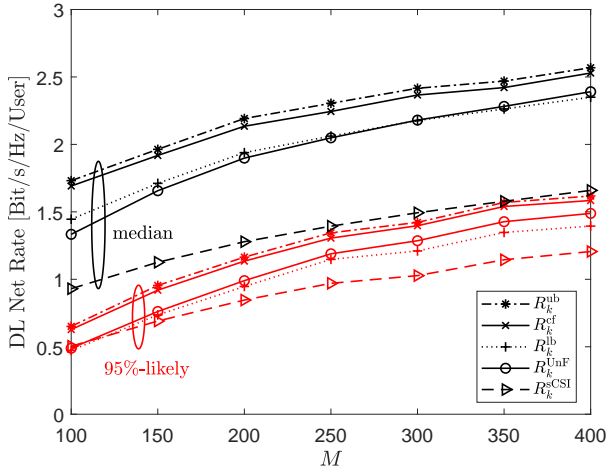} \vspace{-2mm}
	\caption{Downlink net rate against number of APs, for different TDD operations and capacity-bounding techniques. Downlink training introduces relevant gain in terms of median and 95\%-likely net rates. Here, $K=20$, $\tauc = 500$, and $\taup = \taudp = 10$.}  
	\label{fig:fig4}
\end{figure}
Looking at~\Figref{fig:fig4}, we immediately notice the large gain introduced by the downlink beamforming training, in terms of both median and 95\%-likely net rates. For the UE, decoding data relying on the statistical CSI is not efficient due the lack of channel hardening. In such a scenario, even quadrupling the number of antennas does not help much to reduce this gap. Interestingly, we observe that the gap between the proposed approximate achievable rate $R_k^{\mathsf{cf}}$ and its upper bound $R_k^{\mathsf{ub}}$ is negligible. Hence, the optimal power control coefficients that are the output of the optimization problem~\eqref{Problem:Max-Min-DL-rate-closed-form}, based on $R_k^{\mathsf{ub}}$, are good sub-optima for $R_k^{\mathsf{cf}}$. The achievable rate obtained by the using-and-forget capacity-bounding technique is, as expected, a pessimistic lower bound when the channel does not significantly harden. However, we can appreciate how the gap between $R_k^{\mathsf{cf}}$ and $R_k^{\mathsf{UnF}}$ reduces as $M$ grows large. The actual downlink achievable rate for cell-free Massive MIMO with downlink beamforming training, defined in~\eqref{eq:capacity-side-information}, as well as the approximate rate in~\eqref{eq:DLrateApprox} lie certainly between $R_k^{\mathsf{cf}}$ and $R_k^{\mathsf{UnF}}$, and they converge as $M$ increases.   
The achievable downlink rate $R_k^{\mathsf{lb}}$ implies the use of non-coherent detection methods at the UE. We note that $R_k^{\mathsf{lb}}$ almost overlaps $R_k^{\mathsf{UnF}}$, suggesting that downlink training via pilots and non-coherent detection methods can provide comparable rates.

\subsection{Downlink Training Gain under Max-Min Power Control}
We now study the benefits introduced by the downlink beamforming training, evaluating the achievable rates $R_k^{\mathsf{cf}}$, $R_k^{\mathsf{sCSI}}$ at their respective (sub)optimal operation points, by including MMF-PC optimization. We also set $\xi^{\mathsf{DL}}=0.5$, $\tauc = 200$ to support uplink transmission and high-speed UEs, respectively. Recall that the combination of short coherence interval, conjugate beamforming and heavy multiuser interference might lead $R_k^{\mathsf{lb}}$ to take on negative values~\cite{Caire2017a} thus the numerical evaluation of $R_k^{\mathsf{lb}}$ is omitted.
\begin{figure}[!t] \centering
	\includegraphics[width=.91\columnwidth]{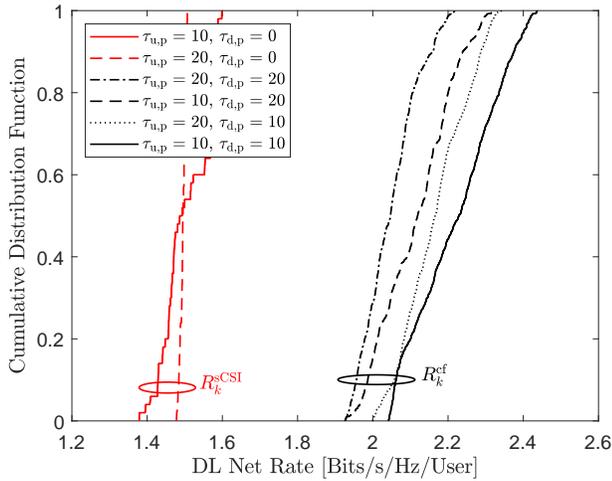} \vspace{-2mm}
	\caption{Per-user downlink net rate, after MMF-PC, for different training duration. Algorithm~\ref{alg:baseline-pilot-assignment} is performed, to satisfy~\eqref{eq:cross-terms-zero}, for the case with $\taudp\!=~\taup\!=~10$, while pilots are assigned at random for all the other cases. Here, $M=200$ and $K = 20$.}  
	\label{fig:fig5}
\end{figure}

\Figref{fig:fig5} shows that downlink beamforming training with MMF-PC can provide performance improvement up to 50\% over the \textit{statistical CSI} case. Therefore, despite the large number of APs and the relative short length of the coherence interval, the downlink training gain is large, justifying the additional pilot overhead and pilot contamination effect\footnote{This significant result was not clear in~\cite{interdonato2016dlpilot}, since the study was limited to mutually orthogonal pilots, and the MMF-PC was not optimized on the achievable downlink rate expression that includes the downlink training.}. 

In this scenario, pilot overhead has relevant impact on the net rate, and its importance is certainly proportional to the length of the coherence interval. For instance, we observe that, among the different configurations of training duration analyzed, $\taudp = \taup = K/2$ provides the best performance, while $\taudp = \taup = K$ is the worst case. This also suggests that the pilot overhead is dominant over the pilot contamination effect. Moreover, looking at the performance of the cases $\taudp = K,~\taup = K/2$, and $\taudp = K/2,~\taup = K$, we clearly deduct that the uplink pilot contamination degrades more the performance than the downlink pilot contamination. The uplink pilot contamination reduces the quality of the channel estimates acquired at the AP, and consequently the accuracy of the beamforming (in such a reciprocity-based system). Hence, the downlink pilot and data transmission are affected by the uplink pilot contamination effect which, at the UE side, sums up to the downlink pilot contamination effect.
\begin{figure}[!t] \centering
	\includegraphics[width=.9\columnwidth]{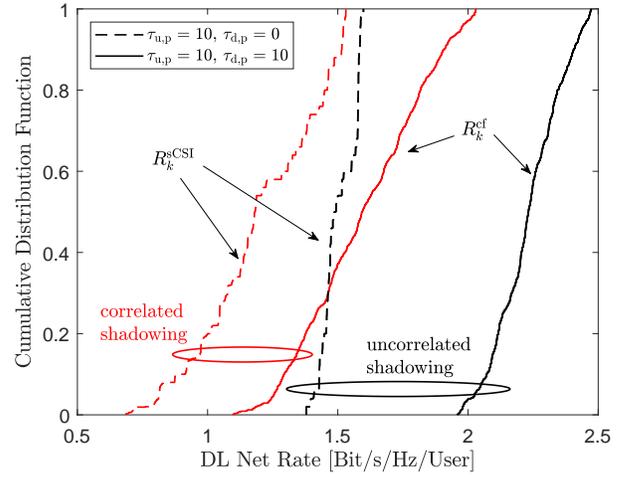} \vspace{-2mm}
	\caption{Per-user downlink net rate with MMF-PC in presence of correlated shadowing. Algorithm~\ref{alg:baseline-pilot-assignment} is performed only for the case with $\taudp\!= \!\taup\!=\!10$, while uplink pilots are assigned at random for the other case. Here, $M=200$, $K=20$.}  
	\label{fig:fig5b}
\end{figure}

In \Figref{fig:fig5b} we analyze the system performance as in~\Figref{fig:fig5} but introducing correlated shadowing in the large-scale fading model. The shadowing correlation model is implemented as in~\cite{Ngo2017b}. We choose the decorrelation distance $d_{\mathrm{decorr}}=0.1$ km which corresponds to a scenario with medium stationarity, and $\delta = 0.5$ to give equal weight to the correlated shadowing at the AP and UE vicinity.
The correlated shadowing drastically reduces the achievable downlink rate. Moreover, without downlink training, we notice that it is more difficult to provide uniform good service, despite the MMF-PC. Importantly, with downlink training, we can still guarantee good rates throughout the network (above 1 bit/s/Hz/user), and increase the performance preserving the same gain (around 50\%) at the 95\%-likely achievable rate, as for the uncorrelated shadowing case study.
   
\subsection{Downlink Training Gain in User-centric Massive MIMO Networks}
In the canonical cell-free Massive MIMO concept, all the APs, geographically distributed over a very wide area, coherently serve all the UEs. However, because of the path loss attenuation, only few APs actually contribute in the transmission to a given UE.  
User-centric networks can be seen as a special case of cell-free Massive MIMO, where a power control policy is applied such that for every UE $k$, only a small number of APs participate in the service of that UE. That is, for each $k$, $\eta_{mk}$ is nonzero only for a small number of APs $m$.
The formation of the user-specific cluster can follow diverse criteria. In~\cite{Ngo2018a}, two AP selection methods are proposed: $(i)$ received-power-based selection, which selects the minimum number of APs that contribute at least $\alpha\%$ of the total received power at the $k$th UE; $(ii)$ largest-large-scale-fading-based selection, which selects the APs with the best channel quality towards UE $k$ as follows
\begin{equation}
\sum\limits_{m=1}^{|\mathcal{A}_k|} \frac{\bar{\beta}_{mk}}{\sum\nolimits_{n=1}^M \beta_{nk}} \geq \alpha\%,
\end{equation}
where $|\mathcal{A}_k|$ is the cardinality of the user-$k$-specific cluster, and $\{\bar{\beta}_{1k}, \ldots, \bar{\beta}_{Mk}\}$ is the set of the large-scale fading coefficients sorted in descending order. 
\begin{figure}[!t] \centering
	\includegraphics[width=.9\columnwidth]{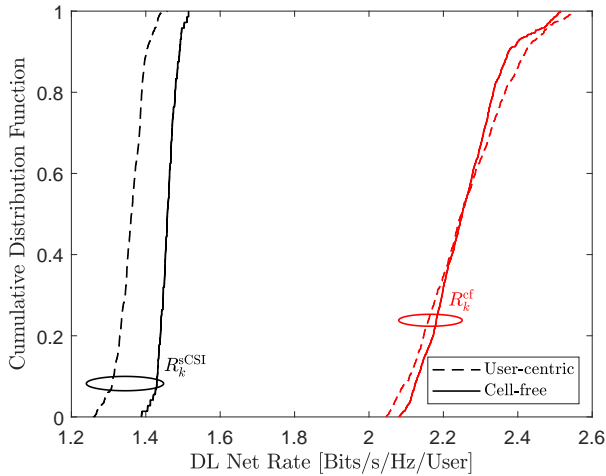} \vspace{-2mm}
	\caption{Per-user downlink net rate, after MMF-PC, for cell-free Massive MIMO and user-centric transmission with and without downlink training. Algorithm~\ref{alg:baseline-pilot-assignment} is performed only for the case with $\taudp = \taup = 10$, while uplink pilots are assigned at random for the other case. Here, $M=200$, $K=20$, and $\alpha=0.95$.}  
	\label{fig:fig6}
\end{figure}

The performance of user-centric network against cell-free Massive MIMO has been recently investigated in~\cite{Buzzi2017b}. Here, we compare the gain introduced by the downlink training in these two different setups, everything else being equal. \Figref{fig:fig6} shows the cdfs of the downlink net rate, for user-centric transmission and cell-free Massive MIMO, both with and without downlink training. In this simulation we used the largest-large-scale-fading-based AP selection method with $\alpha=0.95$, $M=200$, and $K=20$. In such a scenario, about 25 APs on average participate in the transmission to a given UE. Due to the small number of serving APs, the channel hardening degree becomes smaller and, as a consequence, the benefits of including downlink pilots is larger. From~\Figref{fig:fig6}, we can see that downlink training can provide an additional 10\% gain in user-centric network compared to the canonical cell-free Massive MIMO. 

\section{Conclusion}
In this work, we demonstrated that letting the UEs estimate the downlink channel via downlink pilots can improve considerably the achievable downlink rate in cell-free Massive MIMO. We also proposed a  greedy pilot assignment method and devised an approximate max-min fairness power control policy that maximizes the smallest of all the UE rates and ensures uniform quality of service throughout the network. The performance evaluation was based on a closed-form expression for an approximate achievable downlink rate that we derived assuming single-antenna APs, conjugate beamforming and independent Rayleigh fading, and which takes into account channel estimation errors and pilot contamination both at the AP and UE. The proposed achievable rate expression for cell-free Massive MIMO with downlink beamforming training was also compared with expressions obtained from different capacity-bounding techniques. From this study, it turned out that the benefits introduced by the downlink beamforming training are significant, well justifying the  additional pilot overhead and interference from pilot contamination. This holds even for large numbers of APs, when the channel hardening degree is larger, and for relatively short coherence intervals. Especially, downlink training is more beneficial when user-centric transmission is used because of the lesser degree of channel hardening in that case. Potentially, blind channel estimation could be used instead of pilot-based estimation, as in \cite{Ngo2017a} for co-located Massive MIMO, but  further studies are needed in this direction.

\appendix

\subsection{Proof of Proposition~\ref{prop:MMSE-DL-estimation}}
\label{app:MMSE-DL-proof}
To derive the closed-form expression for the downlink channel estimate in~\eqref{eq:a_kk} we compute
\begin{align}
\label{eq:Ea_kk'}
\EX{a_{kk'}} &= \sqrt{\taup\Pup} \sum\nolimits_{m=1}^M \sqrt{\eta_{mk'}} c_{mk'} \nonumber \\
&\qquad \times\EX{ |g_{mk}|^2 \bm{\varphi}\herm_{k} \bm{\varphi}_{k'} + g_{mk} \left( \sum\nolimits^K_{i \neq k} g_{mi} \bm{\varphi}\herm_{k'} \bm{\varphi}_{i} \right)^*} \nonumber \\
&= \bm{\varphi}\herm_{k} \bm{\varphi}_{k'} \sum\limits_{m=1}^M \sqrt{\eta_{mk'}} \gamma_{mk'} \frac{\beta_{mk}}{\beta_{mk'}}, 
\end{align}
where we used~\eqref{eq:uplinkpilotprojection}--\eqref{eq:defGamma}, and the fact that $g_{mk}$ and $g_{mi}$ are uncorrelated, for $k \neq i$, and zero-mean RVs. Note that if $k^\prime = k$ then we directly obtain~\eqref{eq:Ea_kk}. We get~\eqref{eq:Ey_dpk} by inserting~\eqref{eq:Ea_kk'} into $\EX{\check{y}_{\mathrm{dp},k}} = \sqrt{\taudp\Pdp}\sum^K_{k'=1} \bm\psi\herm_{k} \bm\psi_{k'} \EX{a_{kk'}}$, and by imposing~\eqref{eq:pilot-assignment-constraint}. 
Now, we focus on $\cov{\check{y}_{\mathrm{dp},k},{\check{y}_{\mathrm{dp},k}}} = \varx{{\check{y}_{\mathrm{dp},k}}}$. We compute
\begin{align}
\label{eq:Ey_dpk2}
&\EX{|\check{y}_{\mathrm{dp},k}|^2} = \taudp\Pdp \mathsf{E}\Bigg\{\left| \sum\limits^M_{m=1} \sum\limits^K_{k'=1} \bm\psi\herm_{k} \bm\psi_{k'} \sqrt{\eta_{mk'}} c_{mk'} g_{mk} \right. \Bigg. \nonumber \\
&\quad \Bigg. \left. \times \left( \sqrt{\taup\Pup} \sum\limits^K_{i = 1} g_{mi} \bm{\varphi}\herm_{k'} \bm{\varphi}_{i} \right)^* \right|^2\Bigg\} \nonumber \\
&\quad +\taudp\Pdp \EX{\left|\sum\limits^M_{m=1} \sum\limits^K_{k'=1} \bm\psi\herm_{k} \bm\psi_{k'} \sqrt{\eta_{mk'}} g_{mk} c_{mk'} \tilde{w}^*_{mk'} \right|^2} + 1 \nonumber \\
&= \taudp\Pdp \sum\limits^M_{m=1} \sum\limits^K_{k'=1} \eta_{mk'} c^2_{mk'} \beta_{mk} |\bm\psi\herm_{k} \bm\psi_{k'}|^2 \nonumber \\
&\quad + \taudp\Pdp \taup\Pup (\mathcal{T}_1 + \mathcal{T}_2)+1,
\end{align}
where the first equality comes from the fact that $\tilde{w}^*_{mk'}$ is independent of $g_{mi}$, $\forall i$, $k'$ and in the second equality we use the following: $\EX{|X+Y|^2} = \EX{|X|^2} + \EX{|Y|^2}$ if $X$ and $Y$ are two independent RVs and $\EX{X}=0$. Moreover, we defined  
\begin{align}
& \mathcal{T}_1 \triangleq \EX{\left|\sum\limits^M_{m=1} |g_{mk}|^2 \sum\limits^K_{k'=1} \sqrt{\eta_{mk'}} c_{mk'} \bm\psi\herm_{k} \bm\psi_{k'} \bm{\varphi}\herm_{k} \bm{\varphi}_{k'} \right|^2}, 
\end{align}
\begin{align}
& \mathcal{T}_2 \triangleq \EX{\left|\sum\limits^M_{m=1} \sum\limits^K_{k'=1} g_{mk} \sqrt{\eta_{mk'}} c_{mk'} \bm\psi\herm_{k} \bm\psi_{k'} \left( \sum\limits^K_{i \neq k} g_{mi} \bm{\varphi}\herm_{k'} \bm{\varphi}_{i} \right)^*\right|^2}.  
\end{align}
For the sake of brevity, let us define \[b_{mk} \triangleq~\sum\nolimits^K_{k'=1} \sqrt{\eta_{mk'}} c_{mk'} \bm\psi\herm_{k} \bm\psi_{k'} \bm{\varphi}\herm_{k} \bm{\varphi}_{k'}.\] Then
\begin{align}
\label{eq:t1}
\mathcal{T}_1 &= \sum\nolimits^M_{m=1} \EX{|g_{mk}|^4} |b_{mk}|^2 \nonumber \\
&\qquad+ \sum\nolimits^M_{m=1} \sum\nolimits^M_{n \neq m} \EX{|g_{mk}|^2} \EX{|g_{nk}|^2} b_{mk} b_{nk}^* \nonumber \\ 
&= 2 \sum\nolimits^M_{m=1} \beta_{mk}^2 |b_{mk}|^2 + \sum\nolimits^M_{m=1} \sum\nolimits^M_{n \neq m} \beta_{mk} \beta_{nk} b_{mk} b_{nk}^* \nonumber \\
&= \sum\nolimits^M_{m=1} \beta_{mk}^2 |b_{mk}|^2 + \left| \sum\nolimits^M_{m=1} \beta_{mk} b_{mk}\right|^2.
\end{align}
Following the same methodology, the final expression of $\mathcal{T}_2$ is given by
\begin{align}
\label{eq:t2}
\mathcal{T}_2 = \sum\limits^M_{m=1} \sum\limits^K_{k'=1} \sum\limits^K_{i \neq k} \eta_{mk'} c_{mk'}^2 \beta_{mk} \beta_{mi} |\bm\psi\herm_{k} \bm\psi_{k'}|^2  |\bm{\varphi}\herm_{i} \bm{\varphi}_{k'}|^2.
\end{align}
Substitution of \eqref{eq:t1} and \eqref{eq:t2} into \eqref{eq:Ey_dpk2} yields 
\begin{align}
\label{eq:Ey_dpk2-final}
&\EX{|\check{y}_{\mathrm{dp},k}|^2} = 1 +\!\taudp\Pdp\!\sum\limits^M_{m=1}\!\sum\limits^K_{k'=1}\!\eta_{mk'} \gamma_{mk'} \beta_{mk} |\bm\psi\herm_{k} \bm\psi_{k'}|^2 \nonumber \\
&\quad+ \taudp\Pdp \sum\limits^M_{m=1} \sum\limits^K_{k'=1} \sum\limits^K_{i \neq k'} \sqrt{\eta_{mk'}} \sqrt{\eta_{mi}} \gamma_{mk'} \gamma_{mi} \frac{\beta_{mk}^2}{\beta_{mk'} \beta_{mi}} \nonumber \\
&\qquad \times \bm\psi\herm_{k} \bm\psi_{k'} \bm\psi\herm_{i} \bm\psi_{k}  \bm{\varphi}\herm_{k} \bm{\varphi}_{k'} \bm{\varphi}\herm_{i} \bm{\varphi}_{k} \nonumber \\
&\quad + \!\taudp\Pdp \left|\sum\limits^M_{m=1}\!\sum\limits^K_{k'=1}\!\sqrt{\eta_{mk'}} \gamma_{mk'} \frac{\beta_{mk}}{\beta_{mk'}} \bm\psi\herm_{k} \bm\psi_{k'} \bm{\varphi}\herm_{k} \bm{\varphi}_{k'} \right|^2\!.
\end{align}
By inserting~\eqref{eq:Ey_dpk2-final} and~\eqref{eq:Ey_dpk} into \[\varx{\check{y}_{\mathrm{dp},k}} = \EX{\left|\check{y}_{\mathrm{dp},k} \right|^2} - \left|\EX{ \check{y}_{\mathrm{dp},k}}\right|^2,\] and by imposing~\eqref{eq:pilot-assignment-constraint}, we get~\eqref{eq:cov-yk-yk}. Lastly, we compute \[\cov{a_{kk},\check{y}_{\mathrm{dp},k}} = \EX{a_{kk} \check{y}^\ast_{\mathrm{dp},k}} - \EX{a_{kk}}\EX{\check{y}_{\mathrm{dp},k}}^\ast.\] Focusing on the term \[\EX{a_{kk} \check{y}^\ast_{\mathrm{dp},k}} = \sqrt{\taudp\Pdp} \sum\nolimits^K_{k'=1} \bm\psi\herm_{k'} \bm\psi_k \EX{a_{kk} a^\ast_{kk'}},\] we compute
\begin{align}
\label{eq:akkakk'}
& \EX{a_{kk} a_{kk'}^\ast} = \sum\limits^M_{m=1} \sqrt{\eta_{mk}\eta_{mk'}} \EX{|g_{mk}|^2 \hat{g}^*_{mk} \hat{g}^*_{mk'}} \nonumber \\
&\quad+ \sum\limits^M_{m=1}\sum\limits^M_{n \neq m} \sqrt{\eta_{mk}\eta_{nk'}} \EX{{g}_{mk} {g}^*_{nk} \hat{g}^*_{mk} \hat{g}_{nk'} } \nonumber \\
&\stackrel{(a)}{=} \sum\limits^M_{m=1} \sqrt{\eta_{mk}\eta_{mk'}} \frac{\gamma_{mk} \gamma_{mk'}}{\beta_{mk'}} \bigg( \beta_{mk} \bm{\varphi}\herm_{k'} \bm{\varphi}_{k} + \sum\limits^K_{i = 1} \beta_{mi} \bm{\varphi}\herm_{i} \bm{\varphi}_{k} \bm{\varphi}\herm_{k'} \bm{\varphi}_{i} \bigg. \nonumber \\
&\bigg. \qquad  + \frac{\bm{\varphi}\herm_{k'} \bm{\varphi}_{k}}{\taup\Pup} \bigg) + \bm{\varphi}\herm_{k'} \bm{\varphi}_{k} \sum\limits^M_{m=1}\sum\limits^M_{n \neq m} \sqrt{\eta_{mk}\eta_{nk'}}  \gamma_{mk} \gamma_{nk'}\frac{\beta_{nk}}{\beta_{nk'}} \nonumber \\
&=\!\sum^M_{m=1}\!\sqrt{\eta_{mk}\eta_{mk'}} \frac{\gamma_{mk} \gamma_{mk'}}{\beta_{mk'}}\left(\sum\limits^K_{i = 1}\beta_{mi} \bm{\varphi}\herm_{i} \bm{\varphi}_{k} \bm{\varphi}\herm_{k'} \bm{\varphi}_{i}+\frac{\bm{\varphi}\herm_{k'} \bm{\varphi}_{k}}{\taup\Pup}\right) \nonumber \\
&\quad+\!\bm{\varphi}\herm_{k'} \bm{\varphi}_{k}\!\sum\limits^M_{m=1}\!\sum\limits^M_{n = 1}\!\sqrt{\eta_{mk}\eta_{nk'}}\gamma_{mk}\gamma_{nk'}\frac{\beta_{nk}}{\beta_{nk'}},
\end{align}
where $(a)$ follows from the fact that
\begin{align}
&\EX{|g_{mk}|^2 \hat{g}_{mk}^* \hat{g}_{mk'}} = \frac{\gamma_{mk} \gamma_{mk'}}{\beta_{mk'}} \left( \beta_{mk} \bm{\varphi}\herm_{k'} \bm{\varphi}_{k} \right. \nonumber \\
&\left.\quad + \sum\limits^K_{i = 1} \beta_{mi} \bm{\varphi}\herm_{i} \bm{\varphi}_{k} \bm{\varphi}\herm_{k'} \bm{\varphi}_{i} + \frac{\bm{\varphi}\herm_{k'} \bm{\varphi}_{k}}{\taup\Pup} \right)  \\           
&\EX{g_{mk} g^*_{nk} \hat{g}_{mk}^* \hat{g}_{nk'}} = \bm{\varphi}\herm_{k'} \bm{\varphi}_{k} \gamma_{mk} \gamma_{nk'}\frac{\beta_{nk}}{\beta_{nk'}},~\forall n,~n \neq m.
\end{align}
By substituting \eqref{eq:akkakk'}, \eqref{eq:Ea_kk} and \eqref{eq:Ey_dpk} in \[\cov{a_{kk},\check{y}_{\mathrm{dp},k}} = \EX{a_{kk} \check{y}^\ast_{\mathrm{dp},k}} - \EX{a_{kk}}\EX{\check{y}_{\mathrm{dp},k}}^\ast,\] and by imposing~\eqref{eq:pilot-assignment-constraint}, we obtain~\eqref{eq:cov-akk-yk}.

\subsection{Proof of Proposition~\ref{prop:DLrateCF}}
\label{app:DL-rate-proof}
To derive the closed-form expression of achievable downlink rate in~\eqref{eq:DLrateCF}, we firstly compute
\begin{align}
\EX{|\hat{a}_{kk}|^2} &= |\EX{a_{kk}}|^2 + \frac{|\cov{a_{kk},\check{y}_{\mathrm{dp},k}}|^2}{|\cov{\check{y}_{\mathrm{dp},k},\check{y}_{\mathrm{dp},k}}|^2} \varx{\check{y}_{\mathrm{dp},k}} \nonumber \\
&= \left( \sum\limits_{m=1}^M \sqrt{\eta_{mk}} \gamma_{mk} \right)^2+ \kappa_k, 
\end{align}
where in the last equality we use~\eqref{eq:Ea_kk}--\eqref{eq:cov-yk-yk}, and~\eqref{eq:kappa-def}. The mean-square of the channel estimation error is given by $\EX{{|\tilde{a}_{kk}|}^2} = \EX{{|a_{kk}-\hat{a}_{kk}|}^2} = \varsigma_{kk} - \kappa_k$. 

\noindent Lastly, we compute $\EX{|a_{kk'}|^2}$.
\begin{align}
\label{eq:akisq2}
& \EX{|a_{kk'}|^2} = \sum\limits^M_{m=1} \eta_{mk'} \ \EX{|g_{mk} \hat{g}^*_{mk'}|^2 } \nonumber \\
&\quad+ \sum\limits^M_{m=1}\sum\limits^M_{n \neq m} \sqrt{\eta_{mk'}\eta_{nk'}} \ \EX{{g}_{mk} {g}^*_{nk} \hat{g}^*_{mk'} \hat{g}_{nk'} } \nonumber \\
&\stackrel{(b)}{=} \sum\limits^M_{m=1} \eta_{mk'} \gamma_{mk'}^2 \frac{\beta_{mk}}{\beta_{mk'}^2} \left( \beta_{mk} |\bm{\varphi}\herm_{k'} \bm{\varphi}_{k}|^2 + \sum\limits^K_{i = 1} \beta_{mi} |\bm{\varphi}\herm_{k'} \bm{\varphi}_{i}|^2 \right. \nonumber \\ 
&\left.\quad+ \frac{1}{\taup\Pup} \right)\!+\!|\bm{\varphi}\herm_{k'} \bm{\varphi}_{k}|^2 \!\sum\limits^M_{m=1}\sum\limits^M_{n \neq m} \sqrt{\eta_{mk'}\eta_{nk'}} \gamma_{mk'} \gamma_{nk'}\frac{\beta_{mk} \beta_{nk}}{\beta_{mk'} \beta_{nk'}} \nonumber \\
&=\! \sum^M_{m=1}\! \eta_{mk'} \gamma_{mk'}^2\! \frac{\beta_{mk}}{\beta_{mk'}^2}\! \left(\! \beta_{mk}\! |\bm{\varphi}\herm_{k'} \bm{\varphi}_{k}|^2\! +\! \frac{\beta_{mk'}}{\sqrt{\taup\Pup}c_{mk'}}\! \right) \nonumber \\
&\quad + \!\sum^M_{m=1}\!\sum^M_{n \neq m}\! \sqrt{\eta_{mk'}\eta_{nk'}}  \gamma_{mk'} \gamma_{nk'}\!\frac{\beta_{mk} \beta_{nk}}{\beta_{mk'} \beta_{nk'}}\!|\bm{\varphi}\herm_{k'} \bm{\varphi}_{k}|^2\!  \nonumber \\
&= |\bm{\varphi}\herm_{k'} \bm{\varphi}_{k}|^2 \left( \sum\limits_{m=1}^M \sqrt{\eta_{mk'}} \gamma_{mk'} \frac{\beta_{mk}}{\beta_{mk'}} \right)^2 + \sum\limits^M_{m=1} \eta_{mk'} \beta_{mk} \gamma_{mk'},
\end{align} 
where $(b)$ follows from 
\begin{align}
&\EX{|g_{mk} \hat{g}^*_{mk'}|^2} = \gamma_{mk'}^2 \frac{\beta_{mk}}{\beta_{mk'}^2} \left( \beta_{mk} |\bm{\varphi}\herm_{k'} \bm{\varphi}_{k}|^2 \right. \nonumber \\
&\left.\qquad\qquad\qquad\qquad + \sum\limits^K_{i = 1} \beta_{mi} |\bm{\varphi}\herm_{k'} \bm{\varphi}_{i}|^2 + \frac{1}{\taup\Pup} \right), \\
&\EX{{g}_{mk} {g}^*_{nk} \hat{g}^*_{mk'} \hat{g}_{nk'}}\!=\!|\bm{\varphi}\herm_{k'} \bm{\varphi}_{k}|^2  \gamma_{mk'} \gamma_{nk'}\!\frac{\beta_{mk} \beta_{nk}}{\beta_{mk'} \beta_{nk'}},~\forall n,~n \neq m.
\end{align}

\section*{Acknowledgement}
We thank Dr. Emil Bj\"{o}rnson for valuable discussions on pilot assignment strategies,
and one of the anonymous reviewers for constructive criticism that helped improve the paper.

\bibliographystyle{IEEEtran}
\bibliography{IEEEabrv,refs}
\end{document}